%% file: Draft.tex
\documentclass[journal]{IEEEtran}
\usepackage{graphicx}
\usepackage{caption}
\usepackage{subcaption}
\usepackage{tabularx,booktabs}
\usepackage{comment}

\usepackage{adjustbox}
\usepackage{tikz}
\usepackage{pgfplots}
\pgfplotsset{compat=1.18} 
\usepackage{pgfplotstable}
\usetikzlibrary{pgfplots.groupplots}
\usetikzlibrary{pgfplots.dateplot}

\usepackage{float}
\usepackage{color}
\usepackage{multicol}
\usepackage{amsmath,amssymb}
\usepackage{amsfonts}
\usepackage{textcomp}
\usepackage{psfrag}
\usepackage{multimedia}
\usepackage{fancybox}
\usepackage{mathtools}

\usepackage{amsmath}
\DeclareMathOperator*{\argmax}{arg\,max}

\newcommand{\vect}[1]{\ensuremath{\boldsymbol{\mathrm{#1}}}}
\newtheorem{theorem}{Theorem}

\newtheorem{corollary}{Corollary}
\newtheorem{Lemma}{Lemma}

\newcommand {\matr}[2]{\left[\begin{array}{#1}#2\end{array}\right]}

\newcommand {\cmatr}[2]{\left\{\begin{array}{#1}#2\end{array}\right.}

\newcounter{lastnote}

\begin{document}

\title{Economic Model Predictive Control as a Solution to Markov Decision Processes}

\author{Dirk Reinhardt, Akhil S. Anand, Shambhuraj Sawant, S\'ebastien Gros,
\thanks{The authors are with the Department of Engineering Cybernetics, Norwegian University of Science and Technology (NTNU), Trondheim, Norway. E-mail:{\tt\small \{dirk.p.reinhardt, sebastien.gros\}@ntnu.no}}
}

\IEEEtitleabstractindextext{
    \begin{abstract}
        Markov Decision Processes (MDPs) offer a fairly generic and powerful framework to discuss the notion of optimal policies for dynamic systems, in particular when the dynamics are stochastic.
        However, computing the optimal policy of an MDP can be very difficult due to the curse of dimensionality present in solving the underlying Bellman equations.
        Model Predictive Control (MPC) is a very popular technique for building control policies for complex dynamic systems.
        Historically, MPC has focused on constraint satisfaction and steering dynamic systems towards a user-defined reference.
        More recently, Economic MPC was proposed as a computationally tractable way of building optimal policies for dynamic systems. When stochsaticity is present, economic MPC is close to the MDP framework.        In that context, Economic MPC can be construed as atractable heuristic to provide approximate solutions to MDPs.
        However, there is arguably a knowledge gap in the literature regarding these approximate solutions and the conditions for an MPC scheme to achieve closed-loop optimality.
        This chapter aims to clarify this approximation pedagogically, to provide the conditions for MPC to deliver optimal policies, and to explore some of their consequences.

    \end{abstract}

}

\maketitle

\IEEEdisplaynontitleabstractindextext

\IEEEpeerreviewmaketitle

%

\section{Introduction}
\newenvironment{colorsection}[1]{\color{#1}}{}





     The concept of Markov Decision Processes (MDPs) provides a solid framework for sequential decision-making in the context of discrete-time stochastic systems~\cite{puterman2014markov}. In an MDP, the problem is represented by a set of states, a set of actions that can be selected to steer the system, and a stage cost to be minimized, possibly over an infinite horizon. An MDP is often accompanied with a discount factor, which favours the cost in the near future over the long term cost. Various problems, including stochastic planning, learning in robot control, and game theory, have been effectively modeled using MDPs. In fact, MDPs have emerged as the de facto standard formalism for studying sequential decision-making \cite{van2012reinforcement}. They have been applied across diverse fields such as operations research, economics, robotics, and behavioral sciences \cite{feinberg2012handbook}. However, MDPs present important practical challenges--often labeled as the ``curse of dimensionality"--because of the complexity of solving the underlying Bellman Equations when the state-action space is not of very low dimension.

    Economic Model Predictive Control (EMPC) has emerged in the recent literature as an effective tool to deliver optimal control policies for discrete dynamics systems. At each sampling instant, EMPC optimizes a sequence of future control inputs based on predictions of the real system trajectories. These predictions are based on the most recent measurements available on the system and a prediction model~\cite{MPCbook}. The input sequence is designed such that the predicted trajectories minimize the given stage cost, while adhering to the system constraints. Only the first input from this sequence is implemented on the real system. Subsequently, the entire control sequence is updated, thereby creating a control policy. Unlike ``classical" MPC, where stability and constraint satisfaction are in focus, EMPC focuses on the closed-loop performance of the real system at hand, and the stage cost in use is typically of economic nature (e.g. energy usage, time, or financial cost associated with operating a system).
      
    When the system at hand is stochastic, EMPC arguably becomes a practical approach to solve the underlying MDP, in the sense that it aims at delivering a control policy that is close to the optimal one. However, EMPC uses significant approximations when forming that control policy. First, the internal model used in the EMPC scheme to form predictions is often selected as deterministic, and hence delivers de facto inaccurate predictions of the future evolution of the real system. Stochastic models are sometimes used in EMPC, but they are often bound to be fairly coarse. Besides, even if using an accurate stochastic model, EMPC optimizes over sequences of actions rather than over policies and therefore intrinsically differs from a genuine solution for the MDP. Introducing an optimization over policies is sometimes done in MPC (e.g. using scenario trees or via very coarse policy parametrizations), but a strong compromise occurs then between optimality and complexity.
           
    The closed-loop performance of EMPC tends to be impacted by the inaccuracy of the internal model. This holds true regardless of the model type, whether it is derived from first principles, augmented with Machine Learning (ML) tools, or purely data-driven. In the context of stochastic systems and when using a deterministic internal EMPC model, expected-value prediction or max-likelihood models are often used to drive the MPC predictions. However, it is unclear when an internal model conforming to these principles translate into good closed-loop performance, or which conditions the internal model ought to satisfy in order for the EMPC policy to be optimal.
    
    In this chapter, we begin with introducing the fundamentals of discounted MDPs in Sec.~\ref{sec:MDP}. This includes the central Bellman equations and the formulation of constrained MDPs within a feasible domain. We also provide intuitions on the concept of discounting, which are often omitted in the literature and in education. We then discuss the connection between EMPC and MDPs, starting from intuitions and then using a formalism based on the action-value function. We then introduce formal conditions on the EMPC internal model such that the EMPC solves the corresponding MDP, and discuss extensively the consequences of these conditions.

\section{Markov Decision Processes}
\label{sec:MDP}
The problem of producing optimal policies for optimally driving stochastic processes, according to a given economic criterion, is arguably best treated in the context of Markov Decision Processes (MDPs).
The MDP framework considers dynamic systems with underlying state and action sets, and (usually) stochastic state transitions.
More specifically, consider the states $\vect s\in \mathbb S$ and actions $\vect a \in \mathbb A$, with respective sets $\mathbb S$ and $\mathbb A$, and the associated state transition
\begin{align}
    \label{eq:StateTransition}
    \vect s_+ \sim \rho \left(\,.\,|\,\vect s, \vect a\,\right)\,,
\end{align}
defining how a state-action pair $\vect s, \vect a$ yields a new state $\vect s_+$.
In \eqref{eq:StateTransition}, $\varrho$ can be of different nature, depending on the set $\mathbb S$.
For $\mathbb S$ being a discrete countable set, $\rho$ is a conditional probability.
For $\mathbb S\subseteq \mathbb R^n$, where $n$ is the state dimension, $\rho$ can be a conditional probability density or a probability measure.

MDPs have the notion of a control policy at their core.
A control policy is a function $\vect \pi\,:\, \mathbb S\,\mapsto\, \mathbb A$ that attributes an action $\vect a$ to a given state $\vect s$.
A process with dynamics \eqref{eq:StateTransition} in closed-loop with policy $\vect a = \vect \pi\left(\vect s\right)$ becomes a Markov Chain, with the closed-loop transition
\begin{align*}
    \rho^{\vect \pi}\left(\,\vect s_+\,|\, \vect s\,\right) := \rho\left(\,\vect s_+\,|\, \vect s,\vect \pi(\vect s)\,\right)\,.
\end{align*}

\begin{figure}[htbp]
    \centering
    \input{Figures/StatePropagation.tex}
    \caption{Different pathways present in a stochastic process for reaching a state $\vect s_k$ (for $k=10$) starting from initial state $\vect s_0$.}
    \label{fig:Propagation}
\end{figure}
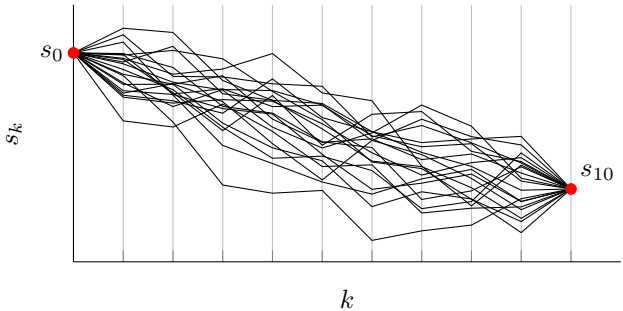

For a given initial condition $\vect s_0$, this Markov Chain yields stochastic state trajectories $\vect s_{0,\ldots,\infty}$ with distribution at any time $k$ given by,
\begin{align*}
    \rho^{\vect \pi}(\,\vect s_k\, |\,\vect s_0\,) = & \int \prod_{i=1}^{k-1} \rho^{\vect \pi}(\vect s_{i}\,|\, \vect s_{i-1})\mathrm d \vect s_{i}\,,
\end{align*}
which compounds the probability of a transition from the initial state $\vect s_0$ at time $0$ to a state $\vect s_k$ at time $k$ via all the possible paths in between, see Fig. \ref{fig:Propagation} for an illustration.

An MDP seeks a policy $\vect \pi$ such that the stochastic closed-loop trajectories minimize a given stage cost $L\,:\, \mathbb S \times  \mathbb A\,\mapsto \, \mathbb R$ in a certain sense.
The most common minimization criterion is arguably the sum of discounted stage costs
\begin{align}
    \label{eq:MDPCost}
    J\left(\vect\pi\right) = \mathbb E\left[\left.\sum_{k=0}^\infty \gamma^k L\left(\vect s_k,\vect a_k\right)\,\right |\, \vect a_k =\vect\pi\left(\vect s_k\right) \right]\,,
\end{align}
for a discount factor $\gamma \in (0,1]$.
In \eqref{eq:MDPCost}, the expectation is taken in the following sense,
\begin{align}
    \label{eq:Individual:E}
     & \mathbb E\left[L\left(\vect s_k,\vect a_k\right)\,|\, \vect a_k = \vect \pi\left(\vect s_k\right)\right] = \\ &\qquad\qquad \int  L\left(\vect s_k,\vect \pi\left(\vect s_k\right)\right)  \rho^{\vect \pi}_k(\,\vect s_k\, |\,\vect s_0\,)\rho_0\left(\vect s_0\right) \mathrm d\vect s_k\mathrm d\vect s_0 \nonumber\,,
\end{align}
where $\rho_0$ is the probability distribution of the initial conditions $\vect s_0$. The solution of an MDP, if it exists, then consists in finding the policy $\vect\pi$ as
\begin{align}
    \label{eq:OptPolicy}
    \min_{\vect\pi}\, J\left(\vect\pi\right)\,.
\end{align}

The cumulative cost criterion \eqref{eq:MDPCost} is the most commonly used in MDPs.
Sec. \ref{sec:MDP:Alternative} discusses alternative criteria.
One motivation for introducing a discounting factor $\gamma$ in \eqref{eq:MDPCost} is to make it bounded and therefore well-defined in the presence of stochastic trajectories.
Indeed, assume that $L\geq 0$ for all state-action pairs.
Then typically, criterion \eqref{eq:MDPCost} with $\gamma = 1$ is unbounded as \eqref{eq:Individual:E} is then greater than zero for all time $k$.
This issue is addressed by selecting $\gamma < 1$, such that costs near in the future are considered more important than the long-term ones.
This geometric weighing ensures that the discounted optimality criterion is bounded under mild conditions.
While discounting is sometimes introduced as a heuristic for ensuring a bounded performance criterion, it has a fairly clear theoretical and practical meaning, which we briefly detail in Sec. \ref{eq:DiscountingMeaning}.

\subsection{Bellman Equations and Value functions}
The solution of an MDP is described through the corresponding Bellman equations, based on the principle of optimality.
The solution to a discounted MDP is described via \cite{bertsekas2019reinforcement} as
\begin{subequations}
    \label{eq:Bellman}
    \begin{align}
        Q^\star\left(\vect s,\vect a\right) & = L\left(\vect s,\vect a\right) + \gamma \mathbb E\left[ V^\star\left(\vect s_+\right)\,|\, \vect s,\vect a \right] \,, \\
        V^\star\left(\vect s\right)         & = \min_{\vect a}\,\, Q^\star\left(\vect s,\vect a\right) \label{eq:Vstar:Bell} \,,                                      \\
        \vect\pi^\star\left(\vect s\right)  & = \mathrm{arg} \min_{\vect a}\,\, Q^\star\left(\vect s,\vect a\right) \,, \label{eq:Pistar:Bell}
    \end{align}
\end{subequations}
where $V^\star$ and $Q^\star$ are labeled value and action-value functions, respectively.
In this discussion, it's important to note the critical role of the action-value function, $Q^\star$, in the context of the Bellman equations. This function is particularly informative as it implicitly defines the other functions through equations \eqref{eq:Vstar:Bell} and \eqref{eq:Pistar:Bell}. While solving the Bellman equations using Dynamic Programming is conceptually straightforward, it presents significant computational challenges. This is primarily due to the "curse of dimensionality," which renders solving equation \eqref{eq:Bellman} nearly impossible except for discrete and low-dimensional state-action spaces. Consequently, most MDPs are only approximately solved to circumvent these difficulties. In this chapter, we aim to explore the extent to which Model Predictive Control (MPC) can be utilized as a tool for solving MDPs, a topic that will be elaborated in Sec. \ref{sec:MPDforMDP}.



%

\subsection{Constraints and MDPs}
\label{sec:ConstrainedMDP}
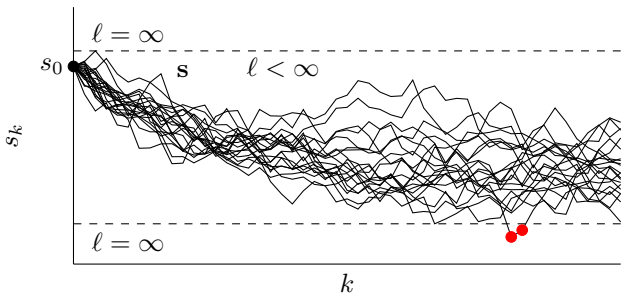
\begin{figure}[htbp]
    \centering
    \input{Figures/MDPConstraints.tex}
    \caption{Markov chains for a policy $\vect \pi$ starting from an initial state $\vect s_0$ along with regions for which the modified stage cost $\ell$ is bounded. (Constraints violations resulting in infinite cost are highlighted in red blobs)}
    \label{fig:Propagation2}
\end{figure}

In some problems, the state of the system is required to remain within a specific subset $\mathbb F$ of the state-action sets $\mathbb S \times \mathbb A$ with unitary probability.
This specification is typically associated to systems subject to strict safety requirements.
It is common and convenient to describe the admissible subset $\mathbb F$ through inequality constraints, typically given as,
\begin{align}
    \mathbb F := \left\{\,\vect s,\vect a\,|\, \vect h\left(\vect s,\vect a\right)\leq 0\,\right\}\,,
\end{align}
for a constraint function $\vect h\,:\,\mathbb S \times \mathbb A\,\mapsto\,\mathbb R^m$.
In the MDP context, such requirements are handled by assigning an infinite penalty to state-action trajectories that leave the admissible set $\mathbb F$.
More specifically, the stage cost $L$ is then modified as,
\begin{align}
    \ell\left(\vect s,\vect a\right) = L\left(\vect s,\vect a\right) + I_{\mathbb F}\left(\vect s,\vect a\right)\,,
\end{align}
where
\begin{align}
    I_{\mathbb F}\left(\vect s,\vect a\right) = \cmatr{ccc}{0 & \text{if} & \left(\vect s,\vect a\right)\in \mathbb F \\ \infty&\text{if}&\left(\vect s,\vect a\right)\notin \mathbb F}
\end{align}
assigns an infinite penalty to state-action pairs leaving the admissible set $\mathbb F$.
Then a policy based on \eqref{eq:OptPolicy} where the optimality criterion $J$ is using the stage cost $\ell$, i.e.,
\begin{align}
    \label{eq:MDPCost:Const}
    J\left(\vect\pi\right) = \mathbb E\left[\left.\sum_{k=0}^\infty \gamma^k \ell\left(\vect s_k,\vect a_k\right)\,\right |\, \vect a_k =\vect\pi\left(\vect s_k\right) \right]\,,
\end{align}
ensures--if it exists--that the system trajectories remain in $\mathbb F$ with probability $1$.
Fig. \ref{fig:Propagation2} illustrates the regions with bounded and unbounded (due to constraint violations) modified stage cost $\ell$ and the resulting Markov chains for a policy $\vect \pi$ and initial state $\vect s_0$.

Clearly, the policy minimizing \eqref{eq:MDPCost:Const} exists only on a set $\mathbb G \subseteq \mathbb F$, implicitly defined as the largest subset of $\mathbb F$ for which there exists a policy $\vect \pi$ that renders $\mathbb G$ invariant for all possible state transitions, more specifically
\begin{align}
    \mathbb P\left[\left(\vect s_+, \vect\pi\left(\vect s_+\right)\right)\in\mathbb G\,|\, \left(\vect s, \vect\pi\left(\vect s\right)\right)\in\mathbb G,\, \vect a = \vect \pi\left(\vect s\right) \right] = 1 \,.
\end{align}

One can verify that the optimal action-value function resulting from a constrained MDP is unbounded outside set $\mathbb G$, i.e.
\begin{align*}
    Q^\star\left(\vect s,\vect a\right) =\infty,\quad \forall \left(\vect s,\vect a\right) \notin \mathbb G
\end{align*}

\subsection{The meaning of discounting} \label{eq:DiscountingMeaning}
Discounting in MDPs is sometimes presented as a heuristic to ensure that the infinite sum of state cost is bounded, and thereby making it a well-posed overall cost to minimize over policies.
This interpretation is unsatisfactory, as it does not give any insights into the role and meaning of the discounting factor.
In this section, we will provide formal insights into that meaning, which can be easily translated into a formal selection of $\gamma$.

To that end, consider an undiscounted MDP where the process has a probabilistic lifetime, with probability $1-\gamma$ of terminating at every discrete time instant, independently of the state trajectory.
The termination results in an interruption of the cost accumulation.
Let us label the lifetime of the process $N\in\mathbb N$.
Assuming that the process is guaranteed to exist at time $k=0$, at every time $k>0$, the cost $L$ has a probability
\begin{align}
    \mathbb P[N\geq k] = 1 - \left(1-\gamma\right)\sum_{i=0}^{k-1} \gamma^i =
    \gamma^{k}
\end{align}
to be incurred.
We can then assemble the overall cost associated with a policy $\vect\pi$ as
\begin{align}
    J\left(\vect\pi\right) & =  \mathbb E\left[\left.\sum_{k=0}^\infty  \mathbb P[N\geq k] \cdot L\left(\vect s_k,\vect a_k\right)\,\right |\vect a_k =\vect\pi\left(\vect s_k\right) \right].
\end{align}
It then becomes trivial to observe that this cost corresponds to the classically discounted MDP setting \eqref{eq:MDPCost}.
Hence the geometric discounting classic in MDPs is a coarse model of a finite process lifetime.

In that context, it becomes interesting to understand what values of $\gamma$ are relevant in engineering applications.
To that end, suppose that a process is sampled at a fairly slow $1\mathrm h$ time rate, and is expected to reach a lifetime of $20$ years with $70\%$ probability.
In that specific case, a $20$ years lifetime corresponds to $175200$ sampling times, and
\begin{align}
    \mathbb P[N\geq 175200] = \gamma^{175200} = 0.7
\end{align}
such that $\gamma = 0.999997964186182$.
Hence long-lasting engineering applications tend to have a discount factor $\gamma$ extremely close to unity.
An alternative interpretation is to view the discount as a coarse model of financial return on investment associated to running a process.
That interpretation yields the same conclusion as the one detailed above.
These observations point to the fact that in engineering applications, discount factors $\gamma$ very close to unity are valid. 

\subsection{Alternative criteria}
\label{sec:AlternativeMDPs}
It is often stated in the MPC literature that MPC schemes aim at optimizing undiscounted optimality criteria.
In the MDP context, i.e. when stochasticity is present in the system, it is then useful to understand properly how such criteria operate.
In this section we briefly cover two important undiscounted optimality criteria for MDPs.
\label{sec:MDP:Alternative}
\subsubsection{Gain Optimality} considers the average cost over time associated to a policy $\vect \pi$. More formally the gain optimality criterion $\bar J$ reads as
\begin{align}
    \label{eq:Gain}
    \bar J\left(\vect\pi\right) =\lim_{M\rightarrow\infty}\frac{1}{M} \mathbb E\left[\left.\sum_{k=0}^M L\left(\vect s_k,\vect a_k\right)\,\right |\, \vect a_k =\vect\pi\left(\vect s_k\right) \right]\,.
\end{align}
A gain-optimal policy is then defined as
\begin{align}
    \label{eq:Gain:Policy}
    \vect\pi^\star = \mathrm{arg}\min_{\vect\pi}\, \bar J\left(\vect\pi\right)\,.
\end{align}
In \eqref{eq:Gain}, the boundedness of the optimality criterion is ensured through the averaging over the horizon.
If $L$ is bounded for all time $k$, then so is $\bar J$.

While sensible, gain-optimal policies are arguably not highly suitable in practice.
The major issue here is that--through averaging--criterion \eqref{eq:Gain} makes the transients of the system irrelevant, focusing only on the long-term optimality of the trajectories.
This has two important consequences: i. the optimal policy \eqref{eq:Gain:Policy} is non-unique, ii. the criterion disregards the short-term costs and focuses only on the long-term ones, hence ignoring the fundamental tenants of economics, whereby gains in the present usually have a higher value than the future ones.
Bias optimality, described next, addresses the first issue.
Discounted optimality, further described in Sec.~\ref{sec:EMPC} and the remainder of the chapter, addresses both.

\subsubsection{Bias Optimality} considers the expected sum of stage costs but offsets each stage cost to ensure the boundedness of the optimality criterion.
The offset is given by the gain-optimality criterion.
More specifically, bias optimality is supported by
\begin{align}
    \label{eq:BO:MDP}
    J^\mathrm{B}\left(\vect\pi\right) = \mathbb E\left[\left.\sum_{k=0}^\infty L\left(\vect s_k,\vect a_k\right) - \bar J\left(\vect\pi\right)\,\right |\, \vect a_k =\vect\pi\left(\vect s_k\right) \right]\,,
\end{align}
and yield the bias-optimal policy
\begin{align}
    \label{eq:Bias:Policy}
    \vect\pi^\star = \mathrm{arg}\min_{\vect\pi}\,  J^\mathrm{B}\left(\vect\pi\right)\,.
\end{align}
Unlike \eqref{eq:Gain:Policy}, under some mild conditions the bias-optimal policy \eqref{eq:Bias:Policy} is unique.
Furthermore, a bias-optimal policy is also a gain-optimal policy, while the opposite is not necessarily true.
Hence, the criterion for bias optimality is more specific than the gain optimal one.
Next, we illustrate the optimal policy obtained for these optimality criteria.


%
%
%

\subsection{Illustrative example} \label{sec:Example:MDP}
\begin{figure}[t]
       \centering
       \def\svgwidth{1\linewidth}
       {\fontsize{10}{10}
           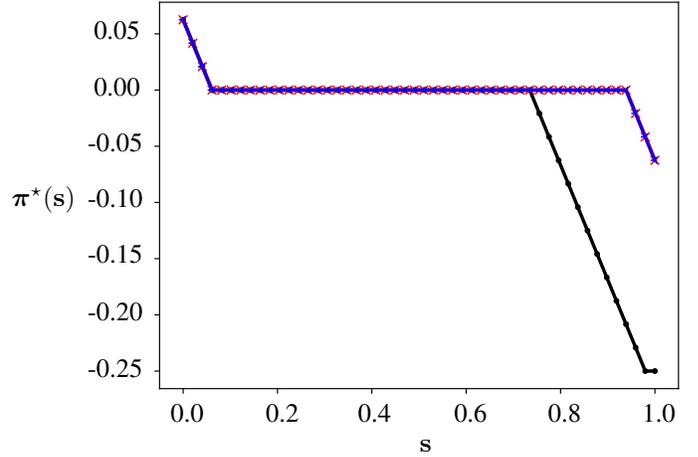}
       \caption{Optimal policies for a) the cumulative cost criterion \eqref{eq:OptPolicy} (black), b) the gain optimality criterion \eqref{eq:Gain:Policy} (red), and c) the bias-optimality criterion \eqref{eq:Bias:Policy} (blue)}
       \label{fig:alternate_policy}
   \end{figure}

Consider a simple example with the scalar dynamics,
\begin{align}
    \label{eq:Dynamics0}
    \vect s_+ = \vect s + \vect a + \vect w,\quad \vect w\sim \mathcal N\left(0,\sigma\right)\,,
\end{align}
with the restriction $\vect s\in [0,1]$ and  $\vect a\in [-0.25,0.25]$ with a discount factor $\gamma = 0.99$.
The additional term $\vect w$ is Gaussian distributed with a standard deviation of $0.05$ and limited to the interval $\vect w\in[-0.05,0.05]$.
Consider the stage cost,
\begin{align}
    \label{eq:PowerCost}
    L\left(\vect s,\vect a\right) = \cmatr{ccc}{\vect a & \text{if} & \vect a \leq 0 \\ 2\vect a&\text{if}&\vect a > 0}\,,
\end{align}
with the addition of high penalties for $\vect s\notin [0,1]$.
Note that this MDP represents a very simplified model of energy storage, where $\vect s$ is the relative stored energy, $\vect a$ is energy purchased or sold (e.g. from/to the power grid), and $\vect w$ the imbalance in local energy consumption-production.
The stage cost \eqref{eq:PowerCost} represents the different costs of buying ($\vect a > 0$) and selling  ($\vect a < 0$) energy.
We observe that for all state $\vect s\in [0,1]$ there is a feasible action $\vect a\in [-0.25,0.25]$ that can keep $\vect s_+\in [0,1]$ regardless of the uncertainty $\vect w$.
Fig. \ref{fig:alternate_policy} shows the optimal policies obtained with the different optimality criteria.

\section{Economic Model Predictive Control}
\label{sec:EMPC}
In general, a basic MPC scheme \cite{rawlings2017model} consists of solving the following optimization problem for a given state $\vect s_k$,
\begin{subequations}
    \label{eq:MPC0}
    \begin{align}
        V^\mathrm{MPC}\left(\vect s_k\right)=\min_{\vect x,\vect u} & \quad T\left(\vect x_{N+1}\right) + \sum_{i=0}^{N} L\left(\vect x_i,\vect u_i\right) \\
        \mathrm{s.t.}                                               & \quad \vect x_{i+1} = \vect f\left(\vect x_i,\vect u_i\right)\label{eq:MPC0:Dyn}     \\
                                                                    & \quad \vect h\left(\vect x_i,\vect u_i\right)\leq 0 \label{eq:MPC0:Const}            \\
                                                                    & \quad \vect x_{0} = \vect s_k,\quad \vect x_{N+1} \in \mathbb T
    \end{align}
\end{subequations}
where $\vect f$ is a model of the system at hand, $\vect h$ is a function representing the input and state limitations that one ought to respect, and $L$ and $T$ are the state cost and terminal cost, respectively.
The terminal set $\mathbb T$ and the terminal cost $T$ are typically used to account for the fact that problem \eqref{eq:MPC0} is solved over a limited horizon $N+1$, while the real system typically runs for a longer time than $N$.
We will assume throughout this chapter that an infinite cost is assigned to an optimization problem of the form \eqref{eq:MPC0} when it is infeasible.

The solution of \eqref{eq:MPC0} delivers a planned sequence of inputs $\vect u^\star_{0,\ldots,N-1}$ and the corresponding state sequence $\vect x^\star_{0,\ldots,N}$ that the system is predicted to follow, according to the selected MPC model $\vect f$.
Because that model is in general inaccurate, this prediction is inexact.
To tackle this issue, problem \eqref{eq:MPC0} is solved again at every discrete time $k$, based on the latest state of the system $\vect s_k$, i.e., the MPC plan is updated at every discrete time, and only the first input of the plan, i.e. $\vect u^\star_{0}$ is effectively applied to the real system.
Therefore, MPC scheme  \eqref{eq:MPC0} produces a policy,
\begin{align}
    \label{eq:MPC:Policy}
    \vect\pi^\mathrm{MPC}\left(\vect s_k\right) = \vect u^\star_0\,,
\end{align}
that assigns for every feasible state $\vect s_k$ of the system a corresponding action implicitly defined by \eqref{eq:MPC0}.
The application of policy \eqref{eq:MPC:Policy} to the real system then yields a state transition from the state-action pair $\vect s_k,\,\vect a_k = \vect\pi^\mathrm{MPC}\left(\vect s_k\right)$ to a next state $\vect s_{k+1}$ and a new action $\vect a_{k+1}$ is evaluated for the next state $\vect s_{k+1}$ from policy \eqref{eq:MPC0}.


In addition to providing a policy $\vect\pi^\mathrm{MPC}$, MPC scheme \eqref{eq:MPC0} also delivers a form of value function $V^\mathrm{MPC}$ through its optimal value.
This observation is at the core of discussing the stability of MPC schemes, see Sec. \ref{sec:Dissipativity}, but it will also play a central role in this work, see Sec. \ref{sec:MPCasMDPmodel}.

\subsection{From tracking MPC to Economic MPC}
Historically, MPC has been arguably focusing on constraint satisfaction.
Indeed, the MPC plan can be easily made to respect state and input limitations by explicitly including them as constraints in \eqref{eq:MPC0:Const}.
The hope is then that the resulting MPC policy \eqref{eq:MPC:Policy} respects these limitations as well.

In that context, the MPC stage cost $L$ is typically designed to bring the system state $\vect s_k$ and action $\vect a_k$ to a chosen reference $\bar{\vect s},\,\bar{\vect a}$, e.g. via simple quadratic penalty functions, i.e.,
\begin{align}\label{eq:QuadCost}
    L\left(\vect x,\vect u\right) = \frac{1}{2}\matr{c}{\vect x-\bar{\vect s} \\\vect u-\bar{\vect a}}^\top W \matr{c}{\vect x-\bar{\vect s}\\\vect u-\bar{\vect a}}\,,
\end{align}
where the state-action pair $\bar{\vect s},\,\bar{\vect a}$ is ideally chosen at a steady-state of the system, and the matrix $W$ is typically positive definite.
The hope with this formulation is that the MPC policy \eqref{eq:MPC:Policy} then steers and stabilizes the real system state and actions to $\bar{\vect s},\,\bar{\vect a}$ while respecting the constraints \eqref{eq:MPC0:Const}, i.e., $\vect h\left(\vect s_k,\vect a_k\right)\leq 0$. In the presence of model inaccuracies and stochasticity in the real system, the MPC policy \eqref{eq:MPC:Policy} may fail at respecting pure state constraints in \eqref{eq:MPC0:Const}. The MPC literature offers a large volume of contributions to address that issue, usually in the form of Robust MPC (RMPC)~\cite{bemporad2007robust} and Stochastic MPC formulations \cite{mesbah2016stochastic}.
\newline

The use of MPC as a way to build policies that seek to obtain the best performances from the real system--in a given sense--rather than steering and stabilising the real system to a given reference, has been the object of a number of publications, usually under the label of Economic MPC.
In that context, the MPC stage cost does not necessarily take the form of a standard quadratic cost such as \eqref{eq:QuadCost}.
Economic stage costs typically linearly relate to the consumption of certain resources within the system, representing the cost of these resources. They can be unbounded from below. For example, in energy-related applications, the economic stage cost often involves the energy usage of the system per sampling interval weighted by the cost of energy.



\subsection{Robust MPC for Stochastic Problems}
Ideally, assuming that the MPC prediction \eqref{eq:MPC0:Const} is perfect, the new state $\vect s_{k+1}$ ought to match the MPC prediction $\vect x_{1}$ in all conditions.
In general, this matching does not occur in practice.
An obvious reason for this mismatch is that the MPC model $\vect f$ is always a simplification of the more complex dynamics governing the real system evolution.
A possibly less obvious reason is that many real systems have, in fact, a stochastic response, whereby a given pair of state and action $\vect s_k,\,\vect a_k$ does not yield a consistent new state $\vect s_{k+1}$, but rather a distribution of new states, conditioned on $\vect s_k,\,\vect a_k$.

The literature on managing model inaccuracies and stochastic state transitions in MPC has a long history, with a strong focus on ensuring that the MPC constraints \eqref{eq:MPC0:Const} are satisfied at all time when the real system operates under policy $\vect\pi^\mathrm{MPC}$.
Arguably the main method to handle constraint satisfaction in the presence of model inaccuracy and stochasticity is RMPC, where a worst-case approach is adopted.
RMPC can take various forms, but most RMPC are arguably approximate solutions to the problem
\begin{align}
    \label{eq:MPC:Robust}
    \min_{\vect \pi}\max_{\vect x\in \mathbb X^N_{\vect \pi}\left(\vect s_k\right)} & \, T\left(\vect x_{N+1}\right) + \sum_{i=0}^{N} \ell\left(\vect x_i,\vect \pi_i\left(\vect x_i\right)\right)\,, 
\end{align}
where set $\mathbb X_{\vect \pi}\left(\vect s_k\right)$ is the set of all possible state trajectories under policy $\vect \pi$ with initial condition $\vect s_k$, formally defined as,
\begin{align}
    \label{eq:RMPC:PossibleTraj}
    \mathbb X^N_{\vect \pi}\left(\vect s_k\right):= & \left\{\vect x_{0,\ldots,N+1}\quad |\quad \vect x_{i+1} \in \sigma(\vect x_i,\vect \pi_i\left(\vect x_i\right))\right. \\&\quad \nonumber\left.\quad\text{and}\quad \vect x_0=\vect s_k \right\}\,,
\end{align}
with $\sigma(\vect x_i,\vect u_i)$ as the support of the probability distribution \eqref{eq:StateTransition} describing the probability of reaching state $ \vect x_{i+1}$ from a given state-action pair $\vect x_i,\vect u_i$.
Note that RMPC usually has a terminal set in which the terminal state $\vect x_{N+1}$ ought to lie for any trajectory $\vect x\in \mathbb X^N_{\vect \pi}\left(\vect s_k\right)$.
The terminal set is in general needed to ensure the recursive feasibility of  \eqref{eq:MPC:Robust}.
Here we assume that the terminal cost $T$ in \eqref{eq:MPC:Robust} is implicitly providing that terminal set, by assigning infinite penalties to solutions $\vect \pi$ that do not ensure recursive feasibility.

Note that similarly to an MDP, the solution of \eqref{eq:MPC:Robust} is delivered by Bellman equations, such that solving \eqref{eq:MPC:Robust} suffers, in general, from the curse of dimensionality.
Hence  \eqref{eq:MPC:Robust} is in general intractable, and usually approximated by replacing the minimization over a generic policy $\vect \pi$ by the more structured and simpler form,
\begin{align}
    \vect \pi_i\left(\vect x_i\right) = \vect v_i - K_i\vect x_i \,,
\end{align}
where $\vect v_i$ are part of the decision variables, and $K_i$ is a sequence of feedback matrices often selected a priori.
Moreover, the conditional support $\sigma$ in \eqref{eq:RMPC:PossibleTraj} is typically approximated as polytopic, with a linear dependency on its argument~\cite{langson2004robust}.
In that context, very computationally effective methods can be developed.


\subsection{From MDP to MPC}
\label{sec:MDPtoMPC}
It can be pedagogically useful to first understand step-by-step how an MPC scheme \eqref{eq:MPC0} approximates an MDP \eqref{eq:MDPCost}.
We will show these steps here. First, we observe that the optimal policy \eqref{eq:OptPolicy} of a classic MDP \eqref{eq:MDPCost} that minimizes \eqref{eq:MDPCost:Const} is the solution of
\begin{align}
    \label{eq:MDP:Short}
    \min_{\vect\pi}\quad \mathbb E\left[\left.\sum_{k=0}^\infty \gamma^k \ell\left(\vect s_k,\vect a_k\right)\,\right |\, \vect a_k =\vect\pi\left(\vect s_k\right) \right]\,.
\end{align}
The first step towards an MPC approximation then consists in transforming \eqref{eq:MDP:Short} into a finite-horizon problem, given by,
\begin{align}
    \label{eq:MDP:Short:Finite}
    \min_{\vect\pi_{0,\ldots,N}}\, \mathbb E\left[\left.\gamma^{N+1}T\left(\vect s_{N+1} \right) + \sum_{k=0}^N \gamma^k \ell\left(\vect s_k,\vect a_k\right)\,\right |\, \vect a_k =\vect\pi_k\left(\vect s_k\right) \right].
\end{align}
We observe that a finite-horizon problem has as solution a sequence of optimal policies $\vect\pi_{0,\ldots,N}^\star$ rather than a single optimal policy $\vect\pi^\star$.
That is because on a finite-horizon problem, the best actions to take when getting close to the problem termination may not be the same as when far away.
In the specific case where
\begin{align}
    \label{eq:ExactTerminalCost}
    T\left(\vect s_{N+1} \right) = V^\star\left(\vect s_{N+1} \right),
\end{align}
then the finite-horizon MDP \eqref{eq:MDP:Short:Finite} coincides with \eqref{eq:MDP:Short} and
\begin{align}
    \vect\pi_{k}^\star = \vect \pi^\star,\qquad k= 0,\ldots,N \,.
\end{align}
It may be useful to stress here that both problems \eqref{eq:MDP:Short} and \eqref{eq:MDP:Short:Finite} suffer from the curse of dimensionality, i.e. both are in general very computationally demanding to solve if the dimensions of states and actions are not small.
A useful alternative to problem \eqref{eq:MDP:Short:Finite} is to observe that for a given initial state $\vect s$, problem \eqref{eq:MDP:Short:Finite} is equivalent to
\begin{align}
    \label{eq:MDP:Short:Finite:FixedFirstInput}
    \min_{\vect a_0,\vect\pi_{1,\ldots,N}}\, \mathbb E\left[\gamma^{N+1}T\left(\vect s_{N+1} \right) + \sum_{k=0}^N \gamma^k \ell\left(\vect s_k,\vect a_k\right)\,\right.\nonumber \\
        \left|\quad \vect s_0=\vect s,\,\vect a_k =\vect\pi\left(\vect s_k\right),\,\, k=1,\ldots,N \phantom{\sum_{k=0}^N}\right]\,,
\end{align}
as the first action $\vect a_0$ is fixed according to the initial state $\vect s_0$, while policies are required for the subsequent time instance due to the stochasticity of the trajectories.
If \eqref{eq:ExactTerminalCost} holds, then 
\begin{align}
    \vect a_0^\star = \vect\pi^\star\left(\vect s\right)
\end{align}
holds for $\vect a_0^\star$ at the solution of \eqref{eq:MDP:Short:Finite:FixedFirstInput} for a given $\vect s$.
If \eqref{eq:ExactTerminalCost} does not hold, then $\vect a_0^\star$ from \eqref{eq:MDP:Short:Finite:FixedFirstInput} is an approximation of the action delivered by the optimal policy $\vect\pi^\star\left(\vect s\right)$.

The second step towards the MPC approximation consists in transforming problem \eqref{eq:MDP:Short:Finite} into a \textit{planning} problem rather than a \textit{policing} problem.
More specifically, consider,
\begin{align}
    \label{eq:MPC:Short:Finite}
    \min_{\vect a_{0,\ldots,N}}\, \mathbb E\left[\, \left.\gamma^{N+1}T\left(\vect s_{N+1} \right) + \sum_{k=0}^N \gamma^k \ell\left(\vect s_k,\vect a_k\right)\,\right |\, \vect s = \vect s_0 \right]\,,
\end{align}
which for a given initial condition $\vect s_0$ fixes an action sequence $\vect a_{0,\ldots,N}$ minimizing the expected cost.
A comparison of \eqref{eq:MPC:Short:Finite} to problem \eqref{eq:MDP:Short:Finite:FixedFirstInput} shows that the only difference is that the optimization over the policy sequence $\vect\pi_{1,\ldots,N}$ in \eqref{eq:MDP:Short:Finite:FixedFirstInput} is now restricted to an action sequence $\vect a_{1,\ldots,N}$ in  \eqref{eq:MPC:Short:Finite}.
A useful interpretation of this second step is that \eqref{eq:MPC:Short:Finite} is a version of \eqref{eq:MDP:Short:Finite:FixedFirstInput} where the search of the optimal policy sequence $\vect\pi^\star_{1,\ldots,N}$ is restricted to a very ``poor" class of policies, namely fixed actions.
Here the optimal first action $\vect a_0^\star$ from the solution of \eqref{eq:MPC:Short:Finite} for $\vect s$ in general does not coincide with the action delivered by the optimal policy $\vect\pi^\star\left(\vect s\right)$.
An exception to that statement is when the MDP trajectories are in fact deterministic.

An important observation here is that \eqref{eq:MPC:Short:Finite} does not suffer from the curse of dimensionality.
Indeed, \eqref{eq:MPC:Short:Finite} can usually be solved as a stochastic program, e.g. by combining Nonlinear Programming and Monte Carlo sampling.
Hence, while the computational demand of solving \eqref{eq:MPC:Short:Finite} accurately can be high, it does not necessarily scale poorly with the dimension of the state and action sets (spaces).
It can finally be useful to observe that stochastic MPC schemes \cite{mesbah2016stochastic} are often aiming at providing approximate solutions to \eqref{eq:MPC:Short:Finite}.

The third step towards a more classic MPC approximation is to replace the stochastic trajectories in \eqref{eq:MPC:Short:Finite} by a deterministic one, hence removing the expected value, and introducing a deterministic model of the stochastic trajectories. This last step yields the discounted MPC scheme,
\begin{subequations}
    \label{eq:MPC:Discounted}
    \begin{align}
        \min_{\vect x,\vect u} & \quad \gamma^{N+1}T\left(\vect x_{N+1}\right) + \sum_{i=0}^{N} \gamma^i L\left(\vect x_i,\vect u_i\right) \\
        \mathrm{s.t.}          & \quad \vect x_{i+1} = \vect f\left(\vect x_i,\vect u_i\right)\,,\label{eq:MPC:Disc:Dyn}                   \\
                               & \quad \vect h\left(\vect x_i,\vect u_i\right)\leq 0\,, \label{eq:MPC:Disc:Const}                          \\
                               & \quad \vect x_{0} = \vect s_k,\quad \vect x_{N+1} \in \mathbb T\,.
    \end{align}
\end{subequations}
Then the first action $\vect a_0^\star=\vect u_0^\star$ delivered by the MPC scheme is a further approximation of the action  $\vect a_0^\star$ delivered by \eqref{eq:MPC:Short:Finite}. We summarize the sequence of approximations as
\begin{align*}
    \underbrace{\eqref{eq:MDP:Short}}_{\text{MDP}}\xrightarrow[]{T\approx V^\star}\eqref{eq:MDP:Short:Finite:FixedFirstInput} \xrightarrow[]{\vect\pi_k\left(\vect s\right)\approx \vect a_k} \eqref{eq:MPC:Short:Finite} \xrightarrow[]{\vect s_{1,\ldots,N} \approx\text{Deterministic}}\underbrace{\eqref{eq:MPC:Discounted}}_{\text{MPC}}\,.
\end{align*}
Finally, the step from the discounted MPC scheme \eqref{eq:MPC:Discounted} to the undiscounted MPC scheme \eqref{eq:MPC0} entails that $\gamma=1$ is used in \eqref{eq:MPC:Discounted}. This can be viewed as a minor approximation when a $\gamma$ very close to 1 is used in the MDP, see Sec. \ref{eq:DiscountingMeaning}. Alternatively, one may consider that the MPC corresponds to an MDP based on undiscounted optimality criteria, see Sec. \ref{sec:AlternativeMDPs}, with the consequence that the system is then assumed to have an unbounded lifetime. A final alternative is to consider a modification of the MPC stage cost $L$ such that the undiscounted MPC corresponds to a discounted MDP, see \cite{kordabad2023equivalence, gros2022economic}.

The main motivation to adopt an MPC scheme \eqref{eq:MPC:Discounted} or \eqref{eq:MPC0} to approximate the optimal policy for an MDP is arguably that solving these MPC problems is tractable (at least to local optimimality), and can often be done in real-time \cite{gros2020linear}.

\section{Dissipativity}
\label{sec:Dissipativity}
In the following, the notion of dissipativity will play a role in understanding MPC as an MDP approximation.
In this section we give some background on dissipativity for both MPC and MDPs.


\subsection{Stability of MPC }
In the field of control, the stability of a policy in closed-loop with a system is a fundamental feature.
Stability entails that, in the absence of disturbances, the closed-loop system will be steered and stabilised to a specific steady state.
In the presence of disturbances or stochasticity, stability can take different forms, such as Input-to-State Stability, stability to a set, as well as more recent results establishing stability in a stochastic sense.

It is worth underlining here that when operating system with economics in focus, stability is not necessarily a crucial feature of the closed-loop system, nor even necessarily a desirable one.
Indeed, some problems with economics in focus achieve their best performances by following a limit-cycle \cite{dockner1991optimality}.
Hence in an economic context, steering the system to a steady-state can be counterproductive.
More generally speaking, if economics is in focus, whichever trajectories minimize the cost or maximize the profit, while respecting the system limitations, is arguably the best solution regardless of whether it yields stability or not.
Nonetheless, economic problems for which it is optimal to steer the trajectories to a steady state constitute a very important class, with important features that we will discuss in this chapter.

It will be useful at this point to recall briefly the theory concluding to the closed-loop stability of a policy stemming from an economic MPC scheme for deterministic dynamics described correctly by the MPC model.
The general form for the nominal stability of MPC is labelled dissipativity theory.
It requires that there exists a so-called storage function $\lambda$ and a class $K_\infty$ function $\kappa$ such that the inequality,
\begin{align}
    \label{eq:Dissipativity0}
    \tilde L\left(\vect s, \vect a\right):= L\left(\vect s, \vect a\right) + \lambda\left(\vect s\right) -  \lambda\left(\vect f\left(\vect s, \vect a\right)\right) \geq \kappa\left(\left|\| \vect s- \bar{\vect s}\right\|\right)\,,
\end{align}
is satisfied for all state-action pairs $\vect s$, $\vect a$, where $\bar{\vect s}$ is the optimal steady state of the system, i.e.,
\begin{subequations}
    \label{eq:optimsteadystate}
    \begin{align}
        \bar{\vect s},\bar{\vect a} = \mathrm{arg}\min_{{\vect s}, {\vect a}} & \quad L\left({\vect s},{\vect a}\right)                           \\
        \mathrm{s.t.}                                                                & \quad  {\vect s} = \vect f\left({\vect s},{\vect a}\right)\,. 
    \end{align}
\end{subequations}
Note that the inequality constraints \eqref{eq:MPC0:Const} have been ignored in the formulation of the optimal state-action pair at steady state \eqref{eq:optimsteadystate}. Indeed, it is an important assumption in dissipativity theory that the state-action pair at steady state belongs to the interior of the feasible domain, such that the inequality constraints are not active at steady state.

For $N$ finite, and some additional conditions required on the terminal cost $T$ of the MPC scheme, inequality \eqref{eq:Dissipativity0} guarantees that if the real system follows the dynamics prescribed by the MPC model $\vect f$, then the closed-loop system is steered to the optimal steady-state $\bar{\vect s},\bar{\vect a}$.
A notable special case of \eqref{eq:Dissipativity0} is the tracking MPC case where the MPC stage cost $L$ readily satisfies \eqref{eq:Dissipativity0} with $\lambda = 0$.
The traditional quadratic stage cost \eqref{eq:QuadCost} is of that nature.

While highly useful, the dissipativity equation \eqref{eq:Dissipativity0} arguably fails at providing a simple intuition as to why satisfying it results in a stabilizing MPC policy.
Fortunately, that intuition is straightforward to build by observing that \eqref{eq:Dissipativity0} guarantees that it is possible to form an MPC scheme equivalent to \eqref{eq:MPC0} with $\ell$ as a stage cost.
More specifically, the MPC scheme
\begin{subequations}
    \label{eq:MPC:Dissipative}
    \begin{align}
        V^\mathrm{D}\left(\vect s_k\right)=\min_{\vect x,\vect u} & \, -\lambda(\vect s_k) + \tilde T\left(\vect x_{N+1}\right) + \sum_{i=0}^{N}  \tilde L\left(\vect x_i,\vect u_i\right) \\
        \mathrm{s.t.}                                             & \quad \vect x_{i+1} = \vect f\left(\vect x_i,\vect u_i\right)\,,                                                       \\
                                                                  & \quad \vect h\left(\vect x_i,\vect u_i\right)\leq 0 \label{eq:MPC:Dissipative:Const} \,,                               \\
                                                                  & \quad \vect x_{0} = \vect s_k,\quad \vect x_{N+1} \in \mathbb T\,,
    \end{align}
\end{subequations}
delivers the same policy and value function as MPC \eqref{eq:MPC0} under the condition that the modified terminal cost is chosen such that $\tilde T(\mathbf{x}_{N+1}) = T(\mathbf{x}_{N+1}) + \lambda(\mathbf{s}_k)$. Here, the initial cost term $\lambda(\mathbf{s}_k)$ summarizes the cost modification \cite{amrit2011economic, gros2019data} in \eqref{eq:Dissipativity0}.

Using the fact that $\tilde L\left(\vect s, \vect a\right) \geq \kappa\left(\left|\| \vect s- \bar{\vect s}\right\|\right)$ is guaranteed by \eqref{eq:Dissipativity0}, one can verify that under some assumptions and for an adequate choice of terminal cost $T$, $V^\mathrm{D}$ is a Lyapunov function for the closed-loop system \cite{diehl2011lyapunov}. Note that dissipativity theory does not apply in the form \eqref{eq:Dissipativity0} to discounted MPC, e.g. in the form \eqref{eq:MPC:Discounted}, but an extension of the dissipativity condition \eqref{eq:Dissipativity0} allows one to apply dissipativity to discounted MPC schemes~\cite{zanon2022new}.

Because \eqref{eq:Dissipativity0} is sufficient but also necessary for the stability of MPC, this observation establishes that the class of stable economic problem in fact belongs to the class of tracking MPC schemes, up to a cost modification.
Equally importantly, if the cost $\tilde L$ is sufficiently differentiable around the steady-state $\bar{\vect s},\bar{\vect a}$, and if that steady state does not activate the MPC inequality constraints, then a stable economic MPC scheme is locally equivalent to an LQR problem~\cite{zanon2016tracking}.
This observation will have interesting consequences further in the text.

\subsection{Dissipativity for MDPs}
An important observation is that the dissipativity theory, which is valid for MPC, does not readily apply to MDPs. The challenge arises because the Lyapunov argument used in MPC, where the MPC value function $V^\mathrm{MPC}$ from \eqref{eq:MPC0} serves as a Lyapunov function for the closed-loop system under certain conditions, is generally not suitable for stochastic dynamics. This issue is elaborated in \cite{gros2022economic}, demonstrating that applying classic dissipativity to stochastic problems necessitates either restricting the focus to stability within a set, akin to Robust MPC, or considering very specific types of stochastic dynamics, such as vanishing perturbations at steady state.

However, dissipativity for MDPs can be discussed in a form nearly identical to \eqref{eq:Dissipativity0} provided that one takes the discussion to a different space than the problem state space.
Let us define $\rho_k^\star$ as the probability measure associated to the MDP state at time $k$ under the optimal policy $\vect\pi^\star$, i.e. the MDP state at time $k$ under the optimal policy is distributed according to
\begin{align}
    \vect s_k\sim \rho_k^\star(.)
\end{align}
and $\rho_0^\star$ is the initial state distribution. Then one can form a dissipativity condition similar to \eqref{eq:Dissipativity0} but in terms of the probability measures $ \rho_k^\star$. More specifically, a dissipativity condition for MDPs reads as,
\begin{align}
    \label{eq:Dissipativity:Stoch}
     & \tilde {\mathcal L}\left[\,\rho_k^\star, \vect \pi^\star(\rho_k^\star)\,\right]:=
    {\mathcal L}\left[\,\rho_k^\star, \vect \pi^\star(\rho_k^\star)\,\right]+ \lambda\left[\,\rho_k^\star\,\right] \nonumber                  \\
     & \qquad\qquad\qquad\qquad - \lambda\left[\,\rho_{k+1}^\star\,\right]  \geq D\left(\, \rho^\star_{k+1}\,||\, \rho_{k}^\star\, \right)\,,
\end{align}
for all $\rho_{k}^\star$, where $ D\left(\, .\,||\, .\, \right)$ is a dissimilarity measure (Kullback-Leibler divergence, Wasserstein distance, total variation distance, etc.). For a ``classic" MDP, ${\mathcal L}$ reads as,
\begin{align}
    {\mathcal L}\left[\,\rho_k^\star, \vect \pi^\star(\rho_k^\star)\,\right] = \mathbb E_{\vect s\sim \rho_k^\star}\left[ L\left(\vect s,\vect \pi^\star(\vect s)\right) \right]\,.
\end{align}
If \eqref{eq:Dissipativity:Stoch} holds and under some additional assumptions on the MDP, similar to the ones needed in classic dissipativity theory~\cite{muller2015necessity}. For $\gamma=1$, \cite{gros2022economic} shows that
\begin{align}
    \lim_{k\rightarrow\infty}D\left(\, \rho^\star_{k}\,||\, {\rho}^\star\, \right) = 0
\end{align}
holds, where $ {\rho}^\star$ is the MDP  optimal steady-state probability measure, solution of
\begin{subequations}
    \label{eq:SteadyStateDist}
    \begin{align}
        {\rho}^\star,\vect\pi^\mathrm{s}=\mathrm{arg} \min_{\rho^\star,\vect\pi} & \quad {\mathcal L}\left[\,\rho^\star, \vect \pi(\rho^\star)\,\right]                                             \\
        \mathrm{s.t.}                                                            & \quad \rho^\star\left(\,.\,\right) = \int \rho[\,.\,|\vect x,\vect\pi(\vect x)]\,\mathrm d\rho^\star(\vect x)\,. 
    \end{align}
\end{subequations}
Similarly to the classic dissipativity theory, this condition can be extended to discounted MDPs, i.e. for $0<\gamma < 1$, see \cite{kordabad2023equivalence}. The notion of MDP dissipativity will be used in the following.

\section{MPC as a solution for MDPs}\label{sec:MPDforMDP}
\label{sec:MPCasMDPmodel}
We now turn to the core of this chapter. In this section, we propose results establishing conditions under which an MPC scheme can deliver the solution of an MDP. Traditionally, MPC has been arguably regarded as a tool for stabilizing systems with state and action limitations to a given reference. In that context, the questions of concern regarding the policy $\vect\pi^\mathrm{MPC}$ resulting from the MPC scheme is constraints satisfaction, usually investigated through recursive feasibility, and stability towards the reference. In this specific context, the connection of MPC to MDPs is of limited interest.

However, the use of ``economic" MPC for operating systems in an economically optimal way has drawn significant attention in the literature for a while, and the benefit of MPC as a tool for improving the economic performance of complex systems has been demonstrated on numerous applications. In that context, the MDP framework is arguably ideal for understanding Economic MPC as a tool for closed-loop performance. Indeed, the MDP framework provides the necessary formal tools to describe the solution to the problem at hand, and an Economic MPC scheme can be seen as a tractable tool to compute an approximation to the optimal policy of an MDP. Hence relating Economic MPC to MDPs and understanding to what extend and in what sense an Economic MPC scheme solves the underlying MDP is of high value. One of the goals of this chapter is to provide such a connection.

An important difficulty in developing this connection is that MPC builds a policy via repeated and possibly inaccurate planning, rather than directly solving the MDP. Indeed
\begin{enumerate}
    \item The MPC model is never a perfect representation of the real system at hand. That imperfection in the MPC predictions does not stem only from possible structural or parametric approximations in the MPC model $\vect f$, which would prevent the prediction to match the real system trajectories. The more fundamental issue, in the context of stochastic systems, is the fact that no deterministic model can reproduce the real system stochasticity. 
    \item The MPC scheme optimizes over a sequence of actions, hence ignoring the closed-loop view adopted in an MDP, where the optimization is over policies. This issue disappears if the true system is deterministic, but is present when it is stochastic.
\end{enumerate}
In addition to these fundamental issues, an MPC scheme typically operates with a finite prediction horizon and without discounting, whereas MDPs are usually formulated as infinite (or long) horizon problems with discounting. To provide further insights into how MPC serves as an approximation of MDPs, we will next elaborate on the approximations made in Sec.\ref{sec:MDPtoMPC}.

\subsection{MPC model}
\label{sec:SYSID}
A crucial step in building an MPC scheme is to select the MPC prediction model $\vect f$ in \eqref{eq:MPC0:Dyn}. Indeed, stage cost $L$ and constraints $\vect h$ in \eqref{eq:MPC0} are typically specified from the problem at hand, while the terminal cost $T$, and terminal set $\mathbb T$ are to be designed in order to compensate for the MPC finite horizon in terms of optimality and recursive feasibility, and are typically based on $L$, $\vect h$ and on the MPC model $\vect f$. Hence, if the MPC scheme is to inherit the cost and constraints $L$, $\vect h$ of the MDP, the MPC model $\vect f$ is arguably the main component in the MPC scheme that can be adjusted such that the MPC solves the MDP. 

\begin{figure}[htbp]
    \centering
    \input{Figures/StateTransitionDensity.tex}
        \caption{Illustration of the difference between state transition models that conform to either the expected value, as defined in \eqref{eq:E:Fitting}, or the maximum likelihood, as given in \eqref{eq:ML:Fitting}, of the MDP's state transition probability density function $\rho$.}
        \label{fig:StatModel}
\end{figure}
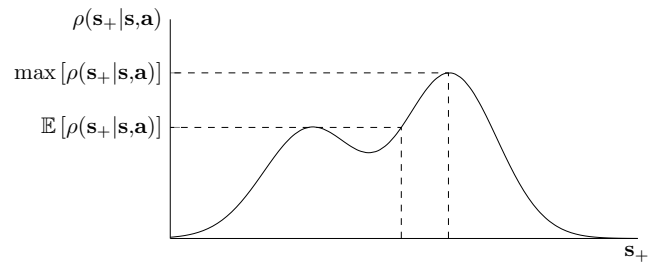

In the context of Economic MPC, the MPC model $\vect f$ is supposed to yield predictions via \eqref{eq:MPC0:Dyn} that are conducive of delivering a policy $\vect\pi^\mathrm{MPC}$ close to the optimal one $\vect\pi^\star$. A common way of building the MPC model is then to seek a function $\vect f$ that simplifies the real system stochastic state transition into one that represents one of its key statistical features. For example, $\vect f$ can represent the expected value of the state transition,
\begin{align}
    \label{eq:E:Fitting}
    \vect f\left(\vect s,\vect a\right) = \mathbb E\left[\,\vect s_+\,|\,\vect s,\vect a\, \right]
\end{align}
or its max-likelihood as
\begin{align}
    \label{eq:ML:Fitting}
    \vect f\left(\vect s,\vect a\right) = \argmax_{\vect s_+} \rho\left(\vect s_+\,|\,\vect s,\vect a \right).
\end{align}
See Fig.~\ref{fig:StatModel} for an illustration.

Approaches \eqref{eq:E:Fitting} and \eqref{eq:ML:Fitting} are arguably the backbone of System Identification \cite{ljung1999system}, and very common for building MPC models. In particular, building models that describe the expected value of the real system state transition as in \eqref{eq:E:Fitting} is the ideal desired outcome when building $\vect f$ from Least-Squares fitting methods.

The hope with building an Economic MPC model $\vect f$ that captures key statistical features of the real system stochastic state transition is that the MPC predictions will be conducive to producing a policy $\vect\pi^\mathrm{MPC}$ close to the optimal one $\vect\pi^\star$. However, the relationship between the lack of accuracy or representativity of the MPC prediction \eqref{eq:MPC0:Dyn} and the degradation of the performance of the MPC policy $\vect\pi^\mathrm{MPC}$ from the optimal policy $\vect\pi^\star$ is complex. In the next section, we will develop tools to address that question. In the remainder of this chapter, we will focus on MPC models seeking to achieve \eqref{eq:E:Fitting}, as they yield the most interesting observations regarding closed-loop optimality.

%
\subsection{MPC as an action-value function}

To investigate the questions raised in this chapter, we need to specify in what sense an MPC scheme solves an MDP. An obvious condition is that the MPC policy $\vect\pi^\mathrm{MPC}$ matches the optimal one $\vect\pi^\star$. However, that condition alone is arguably not fully satisfying. Indeed, the MPC scheme is--by construction--not only a model of the optimal policy but also a model of the MDP value function, as already observed in Sec. \ref{sec:EMPC}. Indeed, for a given state $\vect s_k$ MPC \eqref{eq:MPC0} delivers, alongside a control action $\vect a_k = \vect\pi^\mathrm{MPC}\left(\vect s_k\right)$, a value $V^\mathrm{MPC}\left(\vect s_k\right)$ readily provided by the optimal cost function obtained when solving \eqref{eq:MPC0}  at $\vect s_k$. It is no stretch of intuition to infer that--if MPC \eqref{eq:MPC0} is meant to be an approximate solution of an MDP--$V^\mathrm{MPC}$ is meant to be an approximation of the MDP optimal value function $V^\star$. Indeed, the MPC value function $V^\mathrm{MPC}$ represents the cumulation of the stage cost over the MPC horizon $N$, with the addition of a terminal cost, which is ideally meant to approximate the remainder of the problem up to an infinite horizon. Elaborating on this viewpoint, we consider next the requirements for MPC \eqref{eq:MPC0} to deliver the optimal policy and value function, i.e. that $\vect\pi^\mathrm{MPC}=\vect\pi^\star$ and $V^\mathrm{MPC}=V^\star$.

The approach we propose to adopt here is via specifying how an MPC scheme defines an action-value function $Q^\mathrm{MPC}$, and on whether that action-value function matches the optimal one of the MDP. To that end, consider the following definition:
\begin{subequations}
    \label{eq:MPC:Qmodel}
    \begin{align}
        Q^\mathrm{MPC}\left(\vect s_k,\vect a_k\right):=\min_{\vect x,\vect u} & \quad T\left(\vect x_{N+1}\right) + \sum_{i=0}^{N} L\left(\vect x_i,\vect u_i\right)      \\
        \mathrm{s.t.}                                                          & \quad \vect x_{i+1} = \vect f\left(\vect x_i,\vect u_i\right) \label{eq:MPC:Qmodel:model} \\
                                                                               & \quad \vect h\left(\vect x_i,\vect u_i\right)\leq 0 \label{eq:MPC:Qmodel:Const}           \\
                                                                               & \quad \vect x_{0} = \vect s_k,\quad \vect x_{N+1} \in \mathbb T                           \\
                                                                               & \quad  \vect u_0=\vect a_k. \label{eq:MPC:Qmodel:Aconst}
    \end{align}
\end{subequations}
One can easily verify that $Q^\mathrm{MPC}$ from \eqref{eq:MPC:Qmodel} readily satisfies the fundamental Bellman relationships between optimal action-value functions, value functions and policies, i.e.
\begin{subequations}
    \begin{align}
        V^\mathrm{MPC}\left(\vect s_k\right) & = \phantom{\mathrm{arg}}\min_{\vect a_k}\, Q^\mathrm{MPC}\left(\vect s_k,\vect a_k\right), \\\quad \vect \pi^\mathrm{MPC}\left(\vect s_k\right) &= \mathrm{arg}\min_{\vect a_k}\, Q^\mathrm{MPC}\left(\vect s_k,\vect a_k\right)
    \end{align}
\end{subequations}
hold by construction. Then we will consider that the MPC scheme solves the MDP if the equality:
\begin{align}
    \label{eq:MPC:PerfectModel}
    Q^\mathrm{MPC}\left(\vect s,\vect a\right)=Q^\star\left(\vect s,\vect a\right),\quad \forall\,\vect s,\vect a
\end{align}
holds. If that equality holds, then the MPC delivers an optimal policy, but also the correct value function and action-value function, and therefore provide a full representation of the MDP solution.

Note that one can adopt a less restrictive definition of when an MPC scheme solves the MDP, namely limited to when the MPC policy $\vect\pi^\mathrm{MPC}$ matches the optimal policy $\vect\pi^\star$. In that latter context, the MPC action-value function needs to match the optimal one only in the sense of
\begin{align}
    \label{eq:MPC:PerfectModel:argmin}
    \mathrm{arg}\min_{\vect a}\,Q^\mathrm{MPC}\left(\vect s,\vect a\right)=\mathrm{arg}\min_{\vect a}\,Q^\star\left(\vect s,\vect a\right),\quad \forall\,\vect s.
\end{align}
Note that \eqref{eq:MPC:PerfectModel} implies \eqref{eq:MPC:PerfectModel:argmin}, but the converse is not true, hence making \eqref{eq:MPC:PerfectModel:argmin} less restrictive than \eqref{eq:MPC:PerfectModel}. While less restrictive, definition \eqref{eq:MPC:PerfectModel:argmin} arguably falls short of making the MPC scheme a fully satisfying solution of the MDP. Indeed, while an MPC scheme satisfying \eqref{eq:MPC:PerfectModel} would deliver the optimal policy, it would not predict the expected cost associated to a given state(-action) pair correctly.

One ought to finally observe that \eqref{eq:MPC:PerfectModel} can hold locally in the actions $\vect a$ for all states $\vect s$, and then deliver the optimal policy, the optimal value function and locally the correct action-value function. Hence, arguably, \eqref{eq:MPC:PerfectModel} holding locally in $\vect a$ is--for all practical purposes--as valuable as if it holds globally.

An additional case of interest is if equality \eqref{eq:MPC:PerfectModel} holds up to a constant, i.e. if \eqref{eq:MPC:PerfectModel} is relaxed to
\begin{align}
    \label{eq:MPC:PerfectModel:PlusConstant}
    Q^\mathrm{MPC}\left(\vect s,\vect a\right) + Q_0=Q^\star\left(\vect s,\vect a\right),\quad \forall\,\vect s,\vect a
\end{align}
for some constant $Q_0\in\mathbb R$. Indeed if the MPC scheme yields an action-value function satisfying \eqref{eq:MPC:PerfectModel:PlusConstant}, then it delivers the optimal policy and the optimal value functions up to constant $Q_0$. As we will see later, this case is important to understand the connection between MPC and MDPs. Note that condition \eqref{eq:MPC:PerfectModel:PlusConstant} can be converted to condition \eqref{eq:MPC:PerfectModel} by simply adding constant $V_0$ to the MPC costs \eqref{eq:MPC0} and \eqref{eq:MPC:Qmodel}.

\subsection{Conditions for MPC optimality} \label{sec:OptimalMPC0}
For the reasons detailed in the beginning of this section, there is no reason for MPC \eqref{eq:MPC:Qmodel} to naturally satisfy either \eqref{eq:MPC:PerfectModel}, \eqref{eq:MPC:PerfectModel:argmin}, or \eqref{eq:MPC:PerfectModel:PlusConstant}, and therefore for being a viable solution of the MDP. Satisfying one or several of these conditions requires non-trivial conditions on the MPC scheme, which are not intuitive nor necessarily satisfied when implementing MPC using classic approaches, see Sec. \ref{sec:SYSID}. This section outlines the conditions necessary for an MPC scheme to fulfill \eqref{eq:MPC:PerfectModel:PlusConstant}, thereby delivering an optimal policy, along with a correct representation of the MDP solution. We provide the key condition in the following Theorem.
\vspace{.25cm}
\begin{theorem}
    \label{Th:MPCOptimality}
    If the MPC model $\vect f$ satisfies the equality:
    \begin{align}
        \label{eq:constantmodif}
        \mathbb E\left[V^\star\left(\vect s_+\right)\,|\, \vect s,\vect a\, \right] - V^\star\left(\vect f\left(\vect s,\vect a\right)\right) = V_0
    \end{align}
    and $T = V^{\star}$ is used in MPC \eqref{eq:MPC:Qmodel} then \eqref{eq:MPC:PerfectModel:PlusConstant} holds.
\end{theorem}
\vspace{.5cm}

\begin{IEEEproof}
    Consider the MPC \eqref{eq:MPC:Qmodel} and modified stage cost $\hat{L}(\vect s, \vect a) = L(\vect s, \vect a) + \gamma V_{0} $, and terminal cost, $T(\vect{s}) = V^{\star}(\vect{s})$
    and the associated action value function delivered with a policy $\vect\pi$ as
    \begin{subequations}
        \begin{align}
            \hat{Q}^{\mathrm{MPC}}(\vect s_k, \vect a_k) = & \min_{\vect x, \vect u} \, \gamma^N V^{\star}(\vect{x}_N) + \sum_{i=0}^{N-1} \gamma^i \hat{L}(\vect{x}_i, \vect{u}_i)) \label{eq:MPC:Qmodel2:cost} \\
                                                           & \mathrm{s.t.}\quad \eqref{eq:MPC:Qmodel:model} - \eqref{eq:MPC:Qmodel:Aconst}
        \end{align}
    \end{subequations}
    Using \eqref{eq:constantmodif}, we get,
    \begin{align}
        \hat{Q}^{\mathrm{MPC}}(\vect s_k, \vect a_k) = Q^{\mathrm{MPC}}(\vect s_k, \vect a_k) + gV_0, \quad  g \in \mathbb R
    \end{align}
    where $g$ is some constant.
    \eqref{eq:MPC:Qmodel2:cost} can be further expanded as,
    \begin{align}
        \eqref{eq:MPC:Qmodel2:cost} = & \min_{\vect x, \vect u} \, \gamma^N V^{\star}(\vect{x}_N) +
        \sum_{i=0}^{N-1} \gamma^i \Big(L(\vect{x}_i, \vect u_i) \notag                                         \\
                                      & + \gamma (\mathbb{E} [V^\star(\vect{x}_{i+1})| \vect{x}_i, \vect u_i)]
        - V^\star(\vect f(\vect{x}_i, \vect u_i)))\Big)\,.
    \end{align}
    Using a telescopic sum, we get,
    \begin{equation}\label{eq:Q and A}
        \eqref{eq:MPC:Qmodel2:cost} = \min_{\vect x, \vect u} Q^{\star}(\vect s_k, \vect a_k) + \mathbb{E}\left[\sum_{i=1}^{N-1} \gamma^i A^{\star}\left(\vect{x}_i, \vect u_i\right)\right]
    \end{equation}
    wherein, $A^{\star}(\vect s, \vect a)$ denotes advantage function associated with $V^\star, Q^\star$. It is defined as
    \begin{align}
        A^{\star}(\vect s, \vect a)=\left\{\begin{array}{cc}
                                               Q^{\star}(\vect s, \vect a)-V^{\star}(\vect s) & \text { if }\left|Q^{\star}(\vect s, \vect a)\right|<\infty \\
                                               \infty                                         & \text { otherwise }
                                           \end{array}\right.
    \end{align}
    According to the Bellman equalities \eqref{eq:Bellman} it holds that
    \begin{align}
        \min_{\vect a} A^{\star}(\vect s, \vect a)=0.
    \end{align}
    As a consequence, \eqref{eq:Q and A} reduces to $Q^\star$, which leads to the resulting equation equivalent to \eqref{eq:MPC:PerfectModel:PlusConstant}:
    \begin{equation}
        Q^{\mathrm{MPC}}(\vect s, \vect a) + g V_{0} = Q^{\star}(\vect s, \vect a), \quad  g \in \mathbb R
    \end{equation}
\end{IEEEproof}

The immediate observation on Theorem \ref{Th:MPCOptimality} is that it establishes a condition on the MPC model $\vect f$ to satisfy in order for the MPC scheme to achieve closed-loop optimality. This condition interlaces the optimal value function of the MDP, $V^\star$, and the prediction model $\vect f$ in a non-trivial way. Before further investigating the implications of Theorem  \ref{Th:MPCOptimality}, it is noteworthy that condition \eqref{eq:constantmodif} is \textit{unlike} the conventional criteria used in for constructing a prediction model for MPC. This suggests that an MPC scheme using such conventional criteria to build $\vect f$ does not necessarily yield an optimal policy.  We will further detail that in the next section.

We ought to stress here that our discussions will pertain to models seeking to achieve \eqref{eq:E:Fitting}. In addition to their relevance within MPC, this choice is motivated by the fact that this family of models, in combination with Theorem \ref{Th:MPCOptimality}, produces interesting results regarding the closed-loop optimality of MPC. The subsequent sections will detail this statement.

\subsection{Local optimality of MPC with expected-value model}
\label{sec:LocalOpt}
In this section, we investigate the consequence of Theorem \ref{Th:MPCOptimality} for models that achieve \eqref{eq:E:Fitting}. If model $\vect f$ is constructed such that it achieves \eqref{eq:E:Fitting}, then condition \eqref{eq:constantmodif} becomes:
\begin{align}
    \label{eq:Delta:E}
    \mathbb E\left[V^\star\left(\vect s_+\right)\,|\, \vect s,\vect a\, \right] - V^\star\left(\mathbb E\left[\vect s_+\,|\,\vect s,\vect a \right]\right) = V_0,\quad  \forall\, \vect s,\vect a,
\end{align}
i.e. it requires that the state transition expected value operator $\mathbb E\left[\cdot \right]$ commutes with the optimal value function $V^\star(\cdot)$ up to a constant $V_0$. We can stress here that condition \eqref{eq:Delta:E} is purely related to the properties of the MDP, i.e. the stochastic dynamics \eqref{eq:StateTransition}, and the MDP cost $\ell$ in \eqref{eq:MDPCost}, and not to the ones of the MPC scheme.
In order to facilitate the discussion of \eqref{eq:Delta:E}, let us first consider the minimum attraction set $\mathbb F \subseteq \left\{\,\vect s\in\mathbb S\,|\, V^\star\left(\vect s\right)<\infty\right\}$ associated to dynamics \eqref{eq:StateTransition} and optimal policy $\vect \pi^\star$ implicitly defined as the minimum compact set such that
\begin{align}
    \mathbb P\left[\,\vect s_+ \in \mathbb F\,|\, \vect s \in\mathbb F,\,\vect a\in\mathcal B(\vect\pi^\star(\vect s),r)\right] = 1
\end{align}
holds, where $\mathcal B(\vect\pi^\star(\vect s),r)$ is a ball of radius $r$ centred at $\vect\pi^\star(\vect s)$. Set $\mathbb F$ then defines the set where the MDP trajectories eventually converge, with the addition of a small set of action disturbances. We ought to stress that a compact set $\mathbb F$ does not necessarily exist. Indeed, the compactness assumption is strong, and typically requires that the dynamics \eqref{eq:StateTransition} have a compact support. Besides, set $\mathbb F$ is strongly related to the dissipativity of the MDP. For a dissipative MDP, the closed-loop state distribution under optimal policy $\vect\pi^\star$ converges in the sense of a dissimilarity measure, like KL-divergence, Wasserstein distance, total variation, to the optimal steady-state distribution defined by \eqref{eq:SteadyStateDist}, see \cite{gros2022economic}. In particular, for $r=0$, set $\mathbb F$ is given by the support of $\rho^\star$ if that support is compact. Let us define:
\begin{align}
    \label{eq:Delta:EE}
     & \Delta\left(\vect s,\vect a\right):=\mathbb E\left[V^\star\left(\vect s_+\right)\,|\, \vect s,\vect a\, \right] - V^\star\left(\vect f\left(\vect s,\vect a\right)\right)
\end{align}
with $\vect f\left(\vect s,\vect a\right)$ given by \eqref{eq:E:Fitting}. We observe that satisfying \eqref{eq:Delta:EE} requires that $\Delta$ is constant. We then propose the following Lemma, which provides a useful analysis of $\Delta$.
\vspace{.25cm}
\begin{Lemma}
    \label{Lem:LocalOptimality}
    Let us assume that $V^\star$ is $N\geq 3$ times continuously differentiable on $\mathbb F$. Then for any $ \vect s\in\mathbb F$ and $\vect a\in\mathcal B(\vect\pi^\star(\vect s),r)$:
    \begin{align}
        \label{eq:Delta}
        \Delta\left(\vect s,\vect a\right)= & \frac{1}{2}\mathrm{Tr}\left(\Sigma\left(\vect s,\vect a\right)\nabla^2V^\star\left(\vect f\left(\vect s,\vect a\right)\right)\right) + R\left(\vect s,\vect a\right)
    \end{align}
    where $\Sigma\left(\vect s,\vect a\right)$ is the conditional covariance of the state transition \eqref{eq:StateTransition} and, using the multi-index notation, $R$ is bounded by:
    \begin{align}
        \label{eq:RBound}
        \left|R\left(\vect s,\vect a\right)\right| \leq c\mu_N+ \sum_{|\vect\alpha|=3}^{N-1}\frac{1}{\vect\alpha!}D^{\vect\alpha} V^\star\left(\vect f\left(\vect s,\vect a\right)\right) \mu_{|\vect\alpha|}
    \end{align}
    for some constant $c\geq 0$, with the moments
    \begin{align}
        \mu_k\left(\vect s,\vect a\right) & = \mathbb E\left[\left\|\vect s_+-\vect f\left(\vect s,\vect a\right)\right\|^k\right].
    \end{align}
\end{Lemma}
\begin{IEEEproof}
    Under the proposed assumptions, the optimal value function $V^\star$ admits a $N^\mathrm{th}$-order Taylor expansion on $\mathbb F$ at $\vect f\left(\vect s,\vect a\right)$, i.e.
    \begin{align}
         & V^\star\left(\vect s_+\right) =  V^\star\left(\vect f\left(\vect s,\vect a\right)\right) + \left(\vect s_+-\vect f\left(\vect s,\vect a\right)\right)^\top   \nabla V^\star                      \\
         & +\frac{1}{2} \left(\vect s_+-\vect f\left(\vect s,\vect a\right)\right)^\top \nabla^2 V^\star\left(\vect s_+-\vect f\left(\vect s,\vect a\right)\right) + R\left(\vect s,\vect a\right)\nonumber
    \end{align}
    holds for $\vect s_+\in\mathbb F$, where $ \nabla V^\star$ and $\nabla^2 V^\star$ are evaluated at $\vect f\left(\vect s,\vect a\right)$, and
    $R\left(\vect s,\vect a\right)$ is the Taylor expansion of $V^\star$ at $\vect f\left(\vect s,\vect a\right)$ from order 3 to $N-1$, together with the remainder of order $N$. It follows that \eqref{eq:RBound} holds for constant $c$, given by:
    \begin{align}
        \label{eq:TaylorConstant}
        c & = \frac{1}{N!}\max_{|\vect \alpha|= N,\,\vect s_+\in\mathbb F}\left|D^{\vect\alpha} V^\star\left(\vect s_+\right)\right|
    \end{align}
    By assumption, $\vect s_+\in\mathbb F$ if $\vect s\in\mathbb F$ and $\vect a\in\mathcal B(\vect\pi^\star(\vect s),r)$, which concludes the proof.
\end{IEEEproof}
\vspace{.5cm}
Note that Lemma \ref{Lem:LocalOptimality} does not necessariy require a compact set $\mathbb F$ if $V^\star$ is smooth everywhere and equals its Taylor series. Then Lemma \ref{Lem:LocalOptimality} applies if the moments $\mu_k$ decay fast enough with $k$. We will not detail this case here. Now, we will discuss the implications of Lemma \ref{Lem:LocalOptimality} on the satisfaction of  \eqref{eq:constantmodif}. A fully formal discussion is not provided here for the sake of brevity and will be the object of future publications.
\subsubsection{Deterministic MDPs} are an obvious class readily satisfying \eqref{eq:constantmodif}. In that case $\Sigma = 0$ and $R=0$ such that $\Delta = 0$. While of limited interest, this observation shows that Theorem \ref{Th:MPCOptimality} and Lemma \ref{Lem:LocalOptimality} support the fact that a deterministic MDP can be solved by an MPC having the correct model, terminal cost and terminal set.
\subsubsection{LQR problems} are a second case of interest. If
the optimal value function $V^\star$ is purely quadratic on $\mathbb F$, and $\Sigma$ is constant, then we observe that $R=0$, and that $\nabla^2V^\star$ is constant. As a result,
\begin{equation}
    \Delta\left(\vect s,\vect a\right)=\frac{1}{2}\mathrm{Tr}\left(\Sigma\nabla^2V^\star\right) = V_0 \, ,
\end{equation}
is constant and \eqref{eq:constantmodif} is satisfied. This case corresponds to the class of LQR problems with an additive process noise independent of $\vect s$ and $\vect a$ (i.e. stochastic LQR problems with i.i.d process noise), and generally to the class of MDPs that are equivalent to an LQR on their attraction set $\mathbb F$. For this class of problems, an MPC model satisfying \eqref{eq:E:Fitting} yields an MPC scheme that solves the MDP, up to a simple offset in the value functions. While these are arguably very specific cases, they inform us that an MDP that is locally equivalent to an LQR problem can be locally solved by an MPC scheme with an expectation-based model \eqref{eq:E:Fitting}. This observation is further discussed next.

\subsubsection{Local optimality of MPC} Let us now discuss a more interesting consequence of Lemma \ref{Lem:LocalOptimality}. First, we observe that if $\Sigma\left(\vect s,\vect a\right)$ and $\nabla^2V^\star\left(\vect f\left(\vect s,\vect a\right)\right)$ are bounded and Lipschitz continuous over the domain $\mathbb F$, then the first term in \eqref{eq:Delta} is also Lipschitz continuous. The associated Lipschitz constant is influenced by the degree of variation in the covariance $\Sigma$ of the dynamics \eqref{eq:StateTransition} and the Hessian $\nabla^2V^\star$ The Lipschitz constant is further related to the magnitude of $\Sigma$ and $\nabla^2V^\star$ over $\mathbb F$. Additionally, we observe that $R$ in \eqref{eq:Delta} remains small or nearly constant under conditions where the moments $\mu_{k>2}$ are small or constant, and where higher-order derivatives of $V^\star$ exhibit minimal variation. Consequently, $\Delta$ in \eqref{eq:Delta} tends to be nearly constant across $\mathbb F$, provided that the MDP is sufficiently smooth and either of the following conditions is met:
\begin{enumerate}
    \item $\mathbb F$ is small, i.e. if the MDP is dissipative, and converging to a narrow steady-state distribution $\bar\rho$, or
    \item $\Sigma$, $\nabla^2V^\star$, and $R$ are nearly constant over $\mathbb F$, i.e. if the moments higher than 2 of the state transition $\rho$ in \eqref{eq:StateTransition} are small or if they do not depend strongly on $\vect s,\vect a$ and if the Hessian of the optimal value function does not vary much over $\mathbb F$.
\end{enumerate}

These observations are very technical but point to important practical observations. First, the class of tracking problems with smooth dynamics of compact support and where the stage cost $L$ is quadratic, can fall into the categories listed above. Indeed, tracking problems are typically dissipative, the attraction set $\mathbb F$ can be small, and the value function is smooth where it is bounded. For this class of problems, one can expect that an MPC model based on the expected value model \eqref{eq:E:Fitting} achieves near-optimal closed-loop performances at the steady-state distribution $\bar\rho$.

Similarly, the class of tracking problems arguably extends to ``economic" MDPs with smooth dynamics and generic smooth stage costs $L$, but which achieve dissipativity. Similarly to tracking problems, dissipative MDPs can have a small attraction set $\mathbb F$ and can fall into the categories enumerated above. Therefore, for a smooth dissipative MDP, characterized by steady-state stochastic trajectories converging to a small set, an MPC scheme utilizing the expectation-based model in \eqref{eq:E:Fitting} delivers a policy that is locally closed-loop optimal.

For cases outside these classes of problems, an MPC scheme based on \eqref{eq:E:Fitting} can not guarantee optimality, even locally. Such cases can be, for e.g., MDPs not achieving dissipativity and MDPs based on non-smooth cost functions or non-smooth dynamics. In particular, MDPs subject to exogenous disturbances, such as the strong variation of energy prices in energy-related applications, do not necessarily fall under Theorem \ref{Th:MPCOptimality}. This opens up for discussion on the optimality of MPC schemes in the context of such underlying MDPs.

\subsection{Illustrative examples}

In order to provide an illustration of the arguments made in this section, let us consider two trivial cases based on the example Consider the example MDP in Sec.~\ref{sec:Example:MDP}. The additional term $\vect w$ in the dynamics \eqref{eq:Dynamics0} is Gaussian distributed in a limited interval. Hence for this example $\vect\mu\left(\vect s,\vect a\right)$ and $\Sigma\left(\vect s,\vect a\right)$ in \eqref{eq:Delta} are linear and constant, respectively.

In the following, we will investigate the effect of two different stage costs, using a discount factor $\gamma=0.99$.
\subsubsection{Non-dissipative problem}
\label{sec:BatExample}
\begin{figure}[t]
        \centering
        \def\svgwidth{1\linewidth}
        {\fontsize{10}{10}
            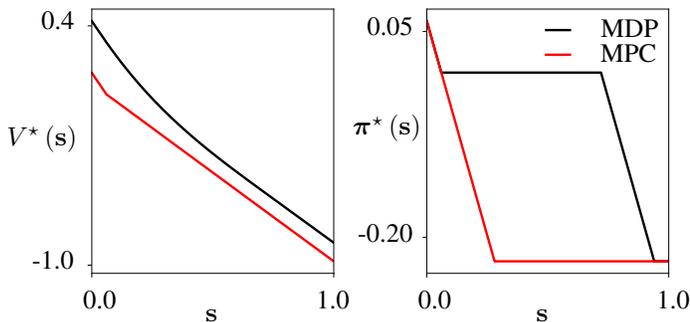}
        \caption{Illustration of the energy storage example in \ref{sec:BatExample}. The optimal value function and policy of the MDP, $V^\star, \pi^\star$ (black) are approximated by the MPC with $V^\mathrm{MPC},\pi^\mathrm{MPC}$ (red). Despite close to linear value functions, the policies differ significantly due to the activation of the lower bound.}
        \label{fig:BatVandPi}
    \end{figure}

First, consider again the example of energy storage with differing prices for buying and selling energy, reflected by the stage cost \eqref{eq:PowerCost} with the additional high penalties for $\vect{s} \notin [0,1]$. We display in Fig. \ref{fig:BatVandPi} the optimal value function $V^\star$ on the feasible domain, optimal policy $\vect \pi^\star$ (in black), and MPC policy $\vect\pi^\mathrm{MPC}$ (in red). Note that the MPC scheme is taking into account the disturbance $\vect w$, to keep $\vect s\in [0,1]$. The cost was, however, calculated based on a nominal model replacing $\vect w$ by its expected value $\mathbb E[\vect w]=0$.

\begin{figure}[t]
    \centering
    \def\svgwidth{1\linewidth}
    {\fontsize{10}{10}
        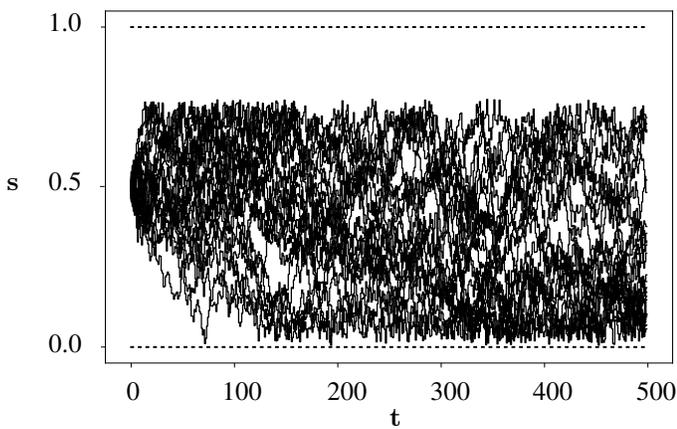}
    \caption{Simulations of the energy storage dynamics \eqref{eq:Dynamics0} of Example~\ref{sec:BatExample}. The problem is not dissipative, and the optimal trajectories do not converge to the neighborhood of a specific steady state.}
    \label{fig:BatSim}
\end{figure}

We observe that even though the optimal value function $V^\star$ appears close to linear, and therefore $\Delta \approx 0$, the MPC policy differs significantly from $\vect \pi^\star$. That is because the optimal trajectories of the problem are not driven to a specific steady state in $[0,1]$, but rather cover a large part of the interval, and activate the lower bound of the state $\vect s=0$, see Fig. \ref{fig:BatSim}.

\subsubsection{Non-smooth value function}
\label{sec:AbsValExample}
Consider the stage cost:
\begin{align}
\label{eq:AbsCost} 
    L\left(\vect s,\vect a\right) = \left|\vect s-\frac{1}{2}\right| +  \left|\vect a\right|
\end{align}
with the addition of high penalties for $\vect s\notin [0,1]$. In this example $\vect w$ is sampled from a normal distribution with standard deviation of $0.1$ and limited to the interval $\vect w\in[-0.25,0.25]$. We observe that for all states $\vect s\in [0,1]$ there is a feasible action $\vect a\in [-0.25,0.25]$ that can keep $\vect s_+\in [0,1]$ regardless of the uncertainty $\vect w$.

\begin{figure}[t]
        \centering
        \def\svgwidth{1\linewidth}
        {\fontsize{10}{10}
            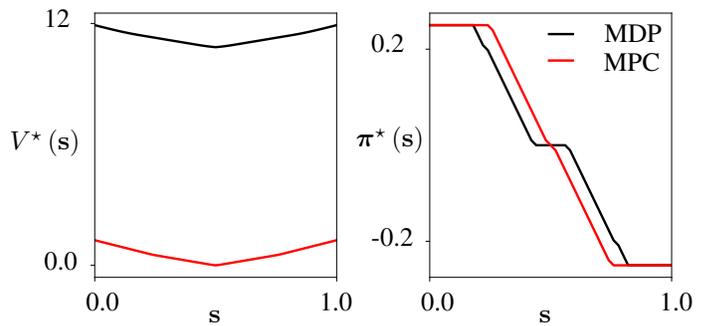}
        \caption{Illustration of the non-smooth value function example in \ref{sec:AbsValExample}. The optimal value function and policy of the MDP, $V^\star, \pi^\star$ (black) are approximated by the MPC with $V^\mathrm{MPC},\pi^\mathrm{MPC}$ (red).}
        \label{fig:AbsVandPi}
    \end{figure}

Fig. \ref{fig:AbsVandPi} shows the optimal value function $V^\star$ on the feasible domain, optimal policy $\vect \pi^\star$ (in black), and MPC policy $\vect\pi^\mathrm{MPC}$ (in red). The MPC scheme was setup in the same way as in Example \ref{sec:BatExample}. We observe that the optimal value function $V^\star$ is non-smooth at $\vect s=\frac{1}{2}$. As a consequence $\Delta$ is not close to zero on this example, see Fig. \ref{fig:AbsDelta}. As a result, the MPC policy differs significantly from the optimal policy $\vect \pi^\star$, see Fig. \ref{fig:AbsVandPi}, right graph.

    \begin{figure}[t]
        \centering
        \def\svgwidth{1\linewidth}
        {\fontsize{10}{10}
            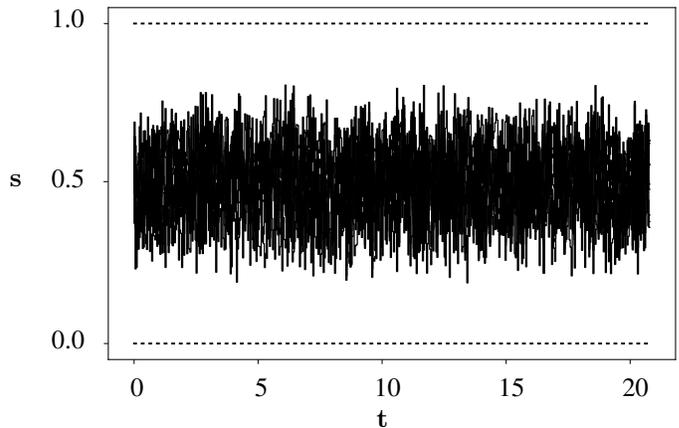}
        \caption{Simulations of Example~\ref{sec:AbsValExample}. The optimal trajectories stay in a region around $\vect s = 0.5$ where the optimal value function is non-smooth.}
        \label{fig:AbsSim}
    \end{figure}




\subsection{Ideal MPC model for performance}
\label{sec:idealmodel}

We finally ought to discuss the implication of Theorem \ref{Th:MPCOptimality} in terms of what it requires from an ``ideal" model for Economic MPC. Indeed, condition \eqref{eq:constantmodif} imposes non-trivial restrictions on the MPC model $\vect f$ which, while potentially difficult to treat in practice, nonetheless provide interesting observations.

\begin{figure}[t]
        \centering
        \def\svgwidth{1\linewidth}
        {\fontsize{10}{10}
            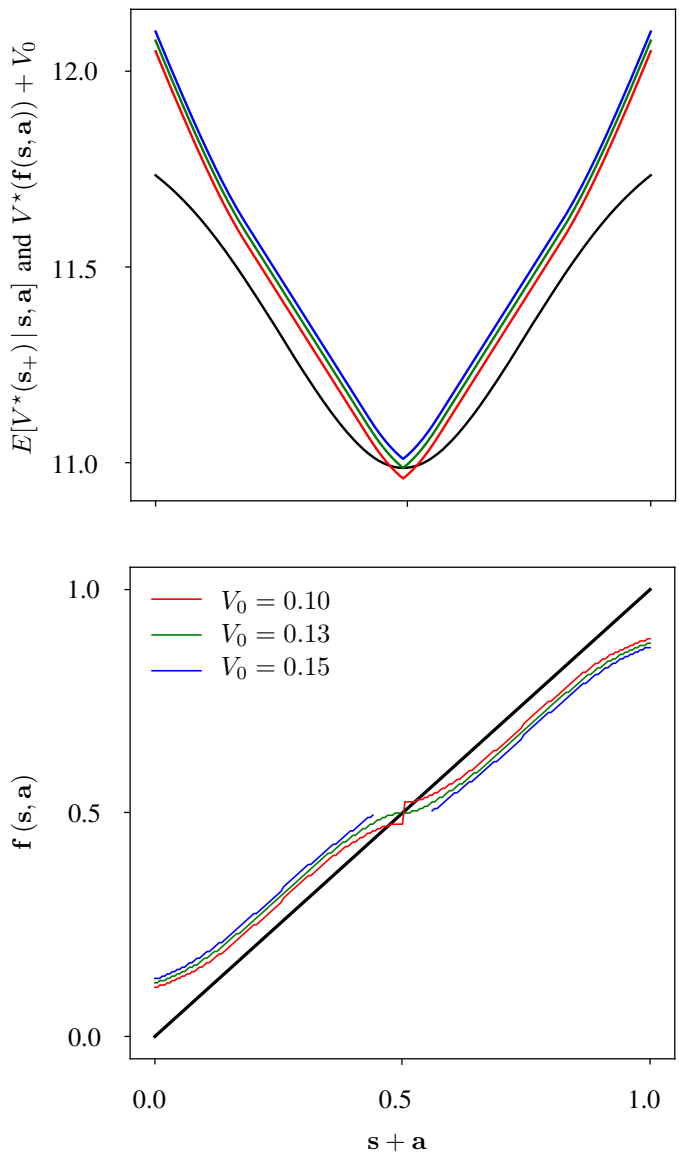}
        \caption{Illustration of the models obtained from Theorem \ref{Th:MPCOptimality} for example \eqref{sec:AbsValExample} and different values of $V_0$. The upper plot shows the resulting value functions (in black for the MDP, in colors for the MPC). The lower graph shows the models in terms of $\vect s+\vect a$. The black line is the expected value of the real state transition, and the colored curves are the optimal models for different values of $V_0$. 
    }
        \label{fig:AbsDelta}
    \end{figure}

First, it ought to be observed that \eqref{eq:constantmodif} does not impose a unique choice of model $\vect f$ per se. Indeed, for a given $\vect s,\vect a$ and for $V_0$ chosen, condition \eqref{eq:constantmodif} only requires $\vect f(\vect s,\vect a)$ to be in the level set:
\begin{align}
\left\{\vect x\quad\mathrm{s.t.}\quad   V^\star\left(\vect x\right) = \mathbb E\left[V^\star\left(\vect s_+\right)\,|\, \vect s,\vect a\, \right]-V_0\right\} 
\end{align}
Hence the selection of an ``ideal" MPC model ought to be accompanied with additional criteria promoting unicity. An additional and fairly obvious criteria to select $\vect f(\vect s,\vect a)$ is to be in the support set of $\rho\left[\vect s_+\,|\, \vect s,\vect a\, \right]$, ideally such that its likelihood is maximized. Hence for a given $V_0$, a meaningful performance-oriented MPC model is one where $\vect f$ is selected according to:
\begin{subequations}
\label{eq:Perf:SYSID}
\begin{align}
\vect f\left(\vect s,\vect a\right) =  &\argmax_{\hat{\vect s}_+}\,\, \rho\left[\, \hat{\vect s}_+\,|\, \vect s,\vect a\, \right] \\
&\mathrm{s.t.}\qquad \mathbb E\left[V^\star\left(\vect s_+\right)\,|\, \vect s,\vect a\, \right] - V^\star\left(\hat{\vect s}_+\right) = V_0
\end{align}
\end{subequations}
for each $\vect s,\vect a$. It is important to note that \eqref{eq:Perf:SYSID} defines the model $\vect f$ for each $\vect s,\vect a$ independently. This implies that the resulting model may lack continuity. While a detailed investigation of this issue is outside the scope of this Chapter, we offer an illustrative trivial example to highlight possible outcomes of \eqref{eq:Perf:SYSID}. To that end, let us revisit the example of Sec.\ref{sec:AbsValExample}, using a non-smooth cost function \eqref{eq:AbsCost} and the linear dynamics \eqref{eq:Dynamics0} with process noise. We then build an MPC model for this problem using \eqref{eq:Perf:SYSID} for different values of $V_0$, and report the results in Fig. \ref{fig:AbsDelta}. 

A few observations are in order here. First, we observe that for some values of $V_0$, \eqref{eq:Perf:SYSID} produces a model $\vect f$ that is not defined everywhere, as indicated by the blue curve in Fig. \ref{fig:AbsDelta}, despite the real system dynamics being fully defined. Second, for some values of $V_0$, \eqref{eq:Perf:SYSID} yields a model $\vect f$ that is not continuous everywhere, as shown by the red curve in Fig. \ref{fig:AbsDelta}, contrary to the real system dynamics being continuous everywhere. Finally, this example includes a specific value of $V_0$ such that the model $\vect f$ from \eqref{eq:Perf:SYSID}  exists everywhere and is continuous, as seen in the green curve in Fig. \ref{fig:AbsDelta}. However, that model is nonlinear even though the real dynamics are linear in expected value. These insights serve as a cautionary note regarding the use of Economic MPC as a way to solve MDPs, because the underlying model that leads the MPC to deliver an optimal policy can have features that are challenging to handle in optimization algorithms. 

\section{Discussion \& Conclusion}

In this Chapter, we explored the application of Economic MPC as an effective method for solving Markov Decision Processes (MDPs). MDPs provide a fairly generic framework for discussing the closed-loop optimization of dynamic systems, particularly under stochastic conditions and with economically driven objectives. We discussed how MPC, possibly discounted, aligns with various optimality criteria in MDPs. While MPC primarily aims at bias optimality, we noted that for many engineering applications, this criterion does not significantly differ from the classical discounted MDP framework. 

We have then discussed the conditions required for an MPC model to generate an optimal policy for a given MDP. These conditions differ from the classic ones used in MPC model construction, such as those involving System Identification techniques. This suggests that when the goal is to enhance the closed-loop performance of an economic MPC scheme, the best model may not come from using classic tools. However, we have also discussed that an exception to this statement is smooth MDPs when the optimal policy consists in steering the system trajectories towards a fixed,  optimal steady state. In that context, an MPC scheme based on traditional modeling techniques can be expected to produce a locally optimal policy in the neighborhood of this steady state. This category of problems includes so-called ``tracking" problems, commonly addressed in MPC, and economic problems achieving dissipativity, provided that there are no strong external disturbances (such as, e.g., strongly varying price signals).  

We ought to conclude this discussion by broadening the perspective offered here in using MPC as a solution to MDPs. The recent literature suggests that when using MPC to solve MDPs, one can benefit from considering the MPC as a model of the MDP solution rather than as an MDP solution based on a model. This approach implies that the MPC model is not the only element that contributes to producing an optimal policy, but that by adjusting the MPC cost and constraints - which may then differ from those of the MDP - matching the optimal policy and value functions between the MPC scheme and the MDP can be achieved. This ``holistic" perspective of the MPC helps to circumvent the difficulties reported in Sec. \ref{sec:idealmodel}. We do not expand on this approach in this Chapter, but it is further detailed in the companion Chapter \textit{final title and number to be inserted.}

\bibliographystyle{IEEEtran}
\bibliography{references}

\end{document}

%% file: Figures/StatePropagation.tex
\begin{tikzpicture}
    \begin{axis}[
            xlabel={$k$},
            ylabel={$s_k$},
            height=5cm,
            width=\columnwidth,
            axis y line*=left,
            axis x line*=bottom,
            ymin=0,
            xmin=0,
            xmax=11,
            ytick={1},      
            yticklabels={$s_0$}, 
            xtick={0,1,...,10},
            xticklabels={},
            xmajorgrids
        ]

        \foreach \i in {0,1,...,19}{
        \addplot +[black, no markers, solid] table[x=k, y=\i,col sep=comma] {Figures/StatePropagationTrajectories.csv};
        }

        \addplot[red, only marks, mark=*] table[x=k, y=0,col sep=comma] {Figures/StatePropagationStartEnd.csv};

        \pgfplotstableread[col sep=comma]{Figures/StatePropagationStartEnd.csv}\datatable
        \pgfplotstablegetelem{1}{k}\of{\datatable}
        \pgfmathsetmacro{\startx}{\pgfplotsretval}

        \pgfplotstablegetelem{1}{0}\of{\datatable}
        \pgfmathsetmacro{\starty}{\pgfplotsretval}
        \node[anchor=south west] at (axis cs:\startx,\starty){$s_{10}$};

    \end{axis}
\end{tikzpicture}

%% file: Figures/MDPConstraints.tex
\begin{tikzpicture}
    \begin{axis}[
            xlabel={$k$},
            ylabel={$s_k$},
            height=5cm,
            width=\columnwidth,
            axis y line*=left,
            axis x line*=bottom,
            ymin=-1,
            ymax= 1.6,
            xmin=0,
            xmax=50,
            ytick={1},      
            yticklabels={$s_0$}, 
            xtick=\empty,  
        ]

        \foreach \i in {0,1,...,19}{
        \addplot +[black, no markers, solid] table[x=k, y=\i,col sep=comma] {Figures/ConstrainedMDP.csv};
        }

        \addplot +[black, no markers, dashed] table[x=k, y=lbx,col sep=comma] {Figures/ConstrainedMDP.csv};
        \addplot +[black, no markers, dashed] table[x=k, y=ubx,col sep=comma] {Figures/ConstrainedMDP.csv};
        \addplot[red, only marks] table[x=k, y=s,col sep=comma] {Figures/ConstrainedMDP_violations.csv};
        \addplot[black, only marks] table[x=k, y=0,col sep=comma] {Figures/ConstrainedMDP_start.csv};

        \node[anchor=south] at (axis cs:10,0.8) {$\mathbf{s}$};
        \node[anchor=south west] at (axis cs:15,0.8) {$\ell<\infty$};

        \pgfplotstableread[col sep=comma]{Figures/ConstrainedMDP.csv}\datatable
        \pgfplotstablegetelem{5}{k}\of{\datatable}
        \pgfmathsetmacro{\startx}{\pgfplotsretval}

        \pgfplotstablegetelem{0}{lbx}\of{\datatable}
        \pgfmathsetmacro{\starty}{\pgfplotsretval}
        \node[anchor=north] at (axis cs:\startx,\starty){$\ell =\infty$};

        \pgfplotstablegetelem{0}{ubx}\of{\datatable}
        \pgfmathsetmacro{\startyy}{\pgfplotsretval}
        \node[anchor=south] at (axis cs:\startx,\startyy){$\ell = \infty$};

    \end{axis}
\end{tikzpicture}

%% file: Figures/DP1D4_b.pdf_tex
\begingroup%
  \makeatletter%
  \providecommand\color[2][]{%
    \errmessage{(Inkscape) Color is used for the text in Inkscape, but the package 'color.sty' is not loaded}%
    \renewcommand\color[2][]{}%
  }%
  \providecommand\transparent[1]{%
    \errmessage{(Inkscape) Transparency is used (non-zero) for the text in Inkscape, but the package 'transparent.sty' is not loaded}%
    \renewcommand\transparent[1]{}%
  }%
  \providecommand\rotatebox[2]{#2}%
  \newcommand*\fsize{\dimexpr\f@size pt\relax}%
  \newcommand*\lineheight[1]{\fontsize{\fsize}{#1\fsize}\selectfont}%
  \ifx\svgwidth\undefined%
    \setlength{\unitlength}{458.47655769bp}%
    \ifx\svgscale\undefined%
      \relax%
    \else%
      \setlength{\unitlength}{\unitlength * \real{\svgscale}}%
    \fi%
  \else%
    \setlength{\unitlength}{\svgwidth}%
  \fi%
  \global\let\svgwidth\undefined%
  \global\let\svgscale\undefined%
  \makeatother%
  \begin{picture}(1,0.6805907)%
    \lineheight{1}%
    \setlength\tabcolsep{0pt}%
    \put(0,0){\includegraphics[width=\unitlength,page=1]{Figures/DP1D4_b.pdf}}%
    \put(0.23684181,0.04908914){\color[rgb]{0,0,0}\makebox(0,0)[lt]{\lineheight{1.25}\smash{\begin{tabular}[t]{l}0.0\end{tabular}}}}%
    \put(0.37986789,0.04908801){\color[rgb]{0,0,0}\makebox(0,0)[lt]{\lineheight{1.25}\smash{\begin{tabular}[t]{l}0.2\end{tabular}}}}%
    \put(0.51835007,0.04908801){\color[rgb]{0,0,0}\makebox(0,0)[lt]{\lineheight{1.25}\smash{\begin{tabular}[t]{l}0.4\end{tabular}}}}%
    \put(0.66233608,0.04908801){\color[rgb]{0,0,0}\makebox(0,0)[lt]{\lineheight{1.25}\smash{\begin{tabular}[t]{l}0.6\end{tabular}}}}%
    \put(0.80350288,0.04908801){\color[rgb]{0,0,0}\makebox(0,0)[lt]{\lineheight{1.25}\smash{\begin{tabular}[t]{l}0.8\end{tabular}}}}%
    \put(0.94493893,0.04908801){\color[rgb]{0,0,0}\makebox(0,0)[lt]{\lineheight{1.25}\smash{\begin{tabular}[t]{l}1.0\end{tabular}}}}%
    \put(0.12164522,0.62487645){\color[rgb]{0,0,0}\makebox(0,0)[lt]{\lineheight{1.25}\smash{\begin{tabular}[t]{l}0.05\end{tabular}}}}%
    \put(0.12164537,0.53960685){\color[rgb]{0,0,0}\makebox(0,0)[lt]{\lineheight{1.25}\smash{\begin{tabular}[t]{l}0.00\end{tabular}}}}%
    \put(0.11224027,0.4564754){\color[rgb]{0,0,0}\makebox(0,0)[lt]{\lineheight{1.25}\smash{\begin{tabular}[t]{l}-0.05\end{tabular}}}}%
    \put(0.11224027,0.37209506){\color[rgb]{0,0,0}\makebox(0,0)[lt]{\lineheight{1.25}\smash{\begin{tabular}[t]{l}-0.10\end{tabular}}}}%
    \put(0.11224027,0.28791067){\color[rgb]{0,0,0}\makebox(0,0)[lt]{\lineheight{1.25}\smash{\begin{tabular}[t]{l}-0.15\end{tabular}}}}%
    \put(0.11224027,0.20220955){\color[rgb]{0,0,0}\makebox(0,0)[lt]{\lineheight{1.25}\smash{\begin{tabular}[t]{l}-0.20\end{tabular}}}}%
    \put(0.11224027,0.11726363){\color[rgb]{0,0,0}\makebox(0,0)[lt]{\lineheight{1.25}\smash{\begin{tabular}[t]{l}-0.25\end{tabular}}}}%
    \put(-0.00181585,0.37353556){\color[rgb]{0,0,0}\makebox(0,0)[lt]{\lineheight{1.25}\smash{\begin{tabular}[t]{l}$\vect \pi^\star (\vect s)$\end{tabular}}}}%
    \put(0.61062277,0.00422921){\color[rgb]{0,0,0}\makebox(0,0)[lt]{\lineheight{1.25}\smash{\begin{tabular}[t]{l}$\vect s$\end{tabular}}}}%
  \end{picture}%
\endgroup%

%% file: Figures/StateTransitionDensity.tex
\adjustbox{max width=\columnwidth}{
    \begin{tikzpicture}

        \newcommand{\ymax}{1.2090451622474927}
        \newcommand{\xmax}{0.595959595959596}
        \newcommand{\ymean}{0.810731363897061}
        \newcommand{\xmean}{0.494949494949495}

        \begin{axis}[
                xlabel={$\mathbf{s}_{+}$},
                xlabel style={at={(ticklabel* cs:1)}, anchor=near ticklabel},
                ylabel={$\rho(\mathbf{s}_{+}\vert\mathbf{s}{,} \mathbf{a})$},
                ylabel style={at={(ticklabel* cs:1)}, anchor=near ticklabel, rotate=-90},
                height=5cm,
                width=\columnwidth,
                axis y line*=left,
                axis x line*=bottom,
                ymin=0.0,
                ymax=1.6,
                xmin=0.0,
                xmax=1.0,
                ytick={\ymean, \ymax},
                yticklabels={$\mathbb{E}\left[\rho(\mathbf{s}_{+}\vert\mathbf{s}{,} \mathbf{a})\right]$, $\mathrm{max}\left[\rho(\mathbf{s}_{+}\vert\mathbf{s}{,} \mathbf{a})\right]$},
                xtick=\empty,
                no markers,
                smooth
            ]
            \addplot+[domain=0:2, samples=100, black]
            {0.8 * exp(-((x - 0.3)^2 / (2 * 0.1^2))) + 1.2 * exp(-((x - 0.6)^2 / (2 * 0.1^2)))};

            \draw[dashed] (axis cs:\xmean,0) -- (axis cs:\xmean,\ymean) -- (axis cs:0.0,\ymean);

            \draw[dashed] (axis cs:\xmax,0) -- (axis cs:\xmax,\ymax) --(axis cs:0.0,\ymax);

        \end{axis}
    \end{tikzpicture}
}

%% file: Figures/Value_pi.pdf_tex
\begingroup%
  \makeatletter%
  \providecommand\color[2][]{%
    \errmessage{(Inkscape) Color is used for the text in Inkscape, but the package 'color.sty' is not loaded}%
    \renewcommand\color[2][]{}%
  }%
  \providecommand\transparent[1]{%
    \errmessage{(Inkscape) Transparency is used (non-zero) for the text in Inkscape, but the package 'transparent.sty' is not loaded}%
    \renewcommand\transparent[1]{}%
  }%
  \providecommand\rotatebox[2]{#2}%
  \newcommand*\fsize{\dimexpr\f@size pt\relax}%
  \newcommand*\lineheight[1]{\fontsize{\fsize}{#1\fsize}\selectfont}%
  \ifx\svgwidth\undefined%
    \setlength{\unitlength}{558.78934605bp}%
    \ifx\svgscale\undefined%
      \relax%
    \else%
      \setlength{\unitlength}{\unitlength * \real{\svgscale}}%
    \fi%
  \else%
    \setlength{\unitlength}{\svgwidth}%
  \fi%
  \global\let\svgwidth\undefined%
  \global\let\svgscale\undefined%
  \makeatother%
  \begin{picture}(1,0.47176827)%
    \lineheight{1}%
    \setlength\tabcolsep{0pt}%
    \put(0,0){\includegraphics[width=\unitlength,page=1]{Figures/Value_pi.pdf}}%
    \put(0.05165125,0.44093036){\color[rgb]{0,0,0}\makebox(0,0)[lt]{\lineheight{1.25}\smash{\begin{tabular}[t]{l}0.4\end{tabular}}}}%
    \put(0.03765153,0.08075724){\color[rgb]{0,0,0}\makebox(0,0)[lt]{\lineheight{1.25}\smash{\begin{tabular}[t]{l}-1.0\end{tabular}}}}%
    \put(0.11453152,0.02082452){\color[rgb]{0,0,0}\makebox(0,0)[lt]{\lineheight{1.25}\smash{\begin{tabular}[t]{l}0.0\end{tabular}}}}%
    \put(0.46461673,0.02082452){\color[rgb]{0,0,0}\makebox(0,0)[lt]{\lineheight{1.25}\smash{\begin{tabular}[t]{l}1.0\end{tabular}}}}%
    \put(0.8911368,0.39129351){\color[rgb]{0,0,0}\makebox(0,0)[lt]{\lineheight{1.25}\smash{\begin{tabular}[t]{l}MPC\end{tabular}}}}%
    \put(0.89181564,0.42885993){\color[rgb]{0,0,0}\makebox(0,0)[lt]{\lineheight{1.25}\smash{\begin{tabular}[t]{l}MDP\end{tabular}}}}%
    \put(0.29830049,0.00262665){\color[rgb]{0,0,0}\makebox(0,0)[lt]{\lineheight{1.25}\smash{\begin{tabular}[t]{l}$\vect{s}$\end{tabular}}}}%
    \put(-0.00012583,0.26623483){\color[rgb]{0,0,0}\makebox(0,0)[lt]{\lineheight{1.25}\smash{\begin{tabular}[t]{l}$V^\star\left(\vect s\right)$\end{tabular}}}}%
    \put(0.61731936,0.02082452){\color[rgb]{0,0,0}\makebox(0,0)[lt]{\lineheight{1.25}\smash{\begin{tabular}[t]{l}0.0\end{tabular}}}}%
    \put(0.97545525,0.02082452){\color[rgb]{0,0,0}\makebox(0,0)[lt]{\lineheight{1.25}\smash{\begin{tabular}[t]{l}1.0\end{tabular}}}}%
    \put(0.79455417,0.00262824){\color[rgb]{0,0,0}\makebox(0,0)[lt]{\lineheight{1.25}\smash{\begin{tabular}[t]{l}$\vect{s}$\end{tabular}}}}%
    \put(0.5404432,0.43230655){\color[rgb]{0,0,0}\makebox(0,0)[lt]{\lineheight{1.25}\smash{\begin{tabular}[t]{l}0.05\end{tabular}}}}%
    \put(0.52644375,0.12191508){\color[rgb]{0,0,0}\makebox(0,0)[lt]{\lineheight{1.25}\smash{\begin{tabular}[t]{l}-0.20\end{tabular}}}}%
    \put(0.51940943,0.28223552){\color[rgb]{0,0,0}\makebox(0,0)[lt]{\lineheight{1.25}\smash{\begin{tabular}[t]{l}$\vect\pi^\star\left(\vect s\right)$\end{tabular}}}}%
  \end{picture}%
\endgroup%

%% file: Figures/sim1.pdf_tex
\begingroup%
  \makeatletter%
  \providecommand\color[2][]{%
    \errmessage{(Inkscape) Color is used for the text in Inkscape, but the package 'color.sty' is not loaded}%
    \renewcommand\color[2][]{}%
  }%
  \providecommand\transparent[1]{%
    \errmessage{(Inkscape) Transparency is used (non-zero) for the text in Inkscape, but the package 'transparent.sty' is not loaded}%
    \renewcommand\transparent[1]{}%
  }%
  \providecommand\rotatebox[2]{#2}%
  \newcommand*\fsize{\dimexpr\f@size pt\relax}%
  \newcommand*\lineheight[1]{\fontsize{\fsize}{#1\fsize}\selectfont}%
  \ifx\svgwidth\undefined%
    \setlength{\unitlength}{524.26814042bp}%
    \ifx\svgscale\undefined%
      \relax%
    \else%
      \setlength{\unitlength}{\unitlength * \real{\svgscale}}%
    \fi%
  \else%
    \setlength{\unitlength}{\svgwidth}%
  \fi%
  \global\let\svgwidth\undefined%
  \global\let\svgscale\undefined%
  \makeatother%
  \begin{picture}(1,0.62638516)%
    \lineheight{1}%
    \setlength\tabcolsep{0pt}%
    \put(0,0){\includegraphics[width=\unitlength,page=1]{Figures/sim1.pdf}}%
    \put(0.06199109,0.59567098){\color[rgb]{0,0,0}\makebox(0,0)[lt]{\lineheight{1.25}\smash{\begin{tabular}[t]{l}1.0\end{tabular}}}}%
    \put(0.06274774,0.1135676){\color[rgb]{0,0,0}\makebox(0,0)[lt]{\lineheight{1.25}\smash{\begin{tabular}[t]{l}0.0\end{tabular}}}}%
    \put(0.06274774,0.35657148){\color[rgb]{0,0,0}\makebox(0,0)[lt]{\lineheight{1.25}\smash{\begin{tabular}[t]{l}0.5\end{tabular}}}}%
    \put(0.17964057,0.04158635){\color[rgb]{0,0,0}\makebox(0,0)[lt]{\lineheight{1.25}\smash{\begin{tabular}[t]{l}0\end{tabular}}}}%
    \put(0.32302215,0.04158412){\color[rgb]{0,0,0}\makebox(0,0)[lt]{\lineheight{1.25}\smash{\begin{tabular}[t]{l}100\end{tabular}}}}%
    \put(0.4809337,0.04158412){\color[rgb]{0,0,0}\makebox(0,0)[lt]{\lineheight{1.25}\smash{\begin{tabular}[t]{l}200\end{tabular}}}}%
    \put(0.63568191,0.04158412){\color[rgb]{0,0,0}\makebox(0,0)[lt]{\lineheight{1.25}\smash{\begin{tabular}[t]{l}300\end{tabular}}}}%
    \put(0.78962964,0.04158412){\color[rgb]{0,0,0}\makebox(0,0)[lt]{\lineheight{1.25}\smash{\begin{tabular}[t]{l}400\end{tabular}}}}%
    \put(0.57423577,0.00279969){\color[rgb]{0,0,0}\makebox(0,0)[lt]{\lineheight{1.25}\smash{\begin{tabular}[t]{l}\vect{t}\end{tabular}}}}%
    \put(0.94483644,0.04158412){\color[rgb]{0,0,0}\makebox(0,0)[lt]{\lineheight{1.25}\smash{\begin{tabular}[t]{l}500\end{tabular}}}}%
    \put(-0,0.35689406){\color[rgb]{0,0,0}\makebox(0,0)[lt]{\lineheight{1.25}\smash{\begin{tabular}[t]{l}\vect{s}\end{tabular}}}}%
  \end{picture}%
\endgroup%

%% file: Figures/Value_pi_2.pdf_tex
\begingroup%
  \makeatletter%
  \providecommand\color[2][]{%
    \errmessage{(Inkscape) Color is used for the text in Inkscape, but the package 'color.sty' is not loaded}%
    \renewcommand\color[2][]{}%
  }%
  \providecommand\transparent[1]{%
    \errmessage{(Inkscape) Transparency is used (non-zero) for the text in Inkscape, but the package 'transparent.sty' is not loaded}%
    \renewcommand\transparent[1]{}%
  }%
  \providecommand\rotatebox[2]{#2}%
  \newcommand*\fsize{\dimexpr\f@size pt\relax}%
  \newcommand*\lineheight[1]{\fontsize{\fsize}{#1\fsize}\selectfont}%
  \ifx\svgwidth\undefined%
    \setlength{\unitlength}{558.80539301bp}%
    \ifx\svgscale\undefined%
      \relax%
    \else%
      \setlength{\unitlength}{\unitlength * \real{\svgscale}}%
    \fi%
  \else%
    \setlength{\unitlength}{\svgwidth}%
  \fi%
  \global\let\svgwidth\undefined%
  \global\let\svgscale\undefined%
  \makeatother%
  \begin{picture}(1,0.46810695)%
    \lineheight{1}%
    \setlength\tabcolsep{0pt}%
    \put(0,0){\includegraphics[width=\unitlength,page=1]{Figures/Value_pi_2.pdf}}%
    \put(0.05094087,0.44696937){\color[rgb]{0,0,0}\makebox(0,0)[lt]{\lineheight{1.25}\smash{\begin{tabular}[t]{l}12\end{tabular}}}}%
    \put(0.05164865,0.08266495){\color[rgb]{0,0,0}\makebox(0,0)[lt]{\lineheight{1.25}\smash{\begin{tabular}[t]{l}0.0\end{tabular}}}}%
    \put(0.54245661,0.4090866){\color[rgb]{0,0,0}\makebox(0,0)[lt]{\lineheight{1.25}\smash{\begin{tabular}[t]{l}0.2\end{tabular}}}}%
    \put(0.54273185,0.11726568){\color[rgb]{0,0,0}\makebox(0,0)[lt]{\lineheight{1.25}\smash{\begin{tabular}[t]{l}-0.2\end{tabular}}}}%
    \put(0.8917857,0.38114811){\color[rgb]{0,0,0}\makebox(0,0)[lt]{\lineheight{1.25}\smash{\begin{tabular}[t]{l}MPC\end{tabular}}}}%
    \put(0.89246453,0.42408198){\color[rgb]{0,0,0}\makebox(0,0)[lt]{\lineheight{1.25}\smash{\begin{tabular}[t]{l}MDP\end{tabular}}}}%
    \put(0.1145569,0.02082391){\color[rgb]{0,0,0}\makebox(0,0)[lt]{\lineheight{1.25}\smash{\begin{tabular}[t]{l}0.0\end{tabular}}}}%
    \put(0.46463207,0.02082391){\color[rgb]{0,0,0}\makebox(0,0)[lt]{\lineheight{1.25}\smash{\begin{tabular}[t]{l}1.0\end{tabular}}}}%
    \put(0.29832061,0.0026266){\color[rgb]{0,0,0}\makebox(0,0)[lt]{\lineheight{1.25}\smash{\begin{tabular}[t]{l}$\vect{s}$\end{tabular}}}}%
    \put(0.61733031,0.02082391){\color[rgb]{0,0,0}\makebox(0,0)[lt]{\lineheight{1.25}\smash{\begin{tabular}[t]{l}0.0\end{tabular}}}}%
    \put(0.97545596,0.02082391){\color[rgb]{0,0,0}\makebox(0,0)[lt]{\lineheight{1.25}\smash{\begin{tabular}[t]{l}1.0\end{tabular}}}}%
    \put(0.79456,0.00262815){\color[rgb]{0,0,0}\makebox(0,0)[lt]{\lineheight{1.25}\smash{\begin{tabular}[t]{l}$\vect{s}$\end{tabular}}}}%
    \put(-0.00012583,0.26245017){\color[rgb]{0,0,0}\makebox(0,0)[lt]{\lineheight{1.25}\smash{\begin{tabular}[t]{l}$V^\star\left(\vect s\right)$\end{tabular}}}}%
    \put(0.51939092,0.26672391){\color[rgb]{0,0,0}\makebox(0,0)[lt]{\lineheight{1.25}\smash{\begin{tabular}[t]{l}$\vect\pi^\star\left(\vect s\right)$\end{tabular}}}}%
  \end{picture}%
\endgroup%

%% file: Figures/sim2.pdf_tex
\begingroup%
  \makeatletter%
  \providecommand\color[2][]{%
    \errmessage{(Inkscape) Color is used for the text in Inkscape, but the package 'color.sty' is not loaded}%
    \renewcommand\color[2][]{}%
  }%
  \providecommand\transparent[1]{%
    \errmessage{(Inkscape) Transparency is used (non-zero) for the text in Inkscape, but the package 'transparent.sty' is not loaded}%
    \renewcommand\transparent[1]{}%
  }%
  \providecommand\rotatebox[2]{#2}%
  \newcommand*\fsize{\dimexpr\f@size pt\relax}%
  \newcommand*\lineheight[1]{\fontsize{\fsize}{#1\fsize}\selectfont}%
  \ifx\svgwidth\undefined%
    \setlength{\unitlength}{524.2672321bp}%
    \ifx\svgscale\undefined%
      \relax%
    \else%
      \setlength{\unitlength}{\unitlength * \real{\svgscale}}%
    \fi%
  \else%
    \setlength{\unitlength}{\svgwidth}%
  \fi%
  \global\let\svgwidth\undefined%
  \global\let\svgscale\undefined%
  \makeatother%
  \begin{picture}(1,0.6328128)%
    \lineheight{1}%
    \setlength\tabcolsep{0pt}%
    \put(0,0){\includegraphics[width=\unitlength,page=1]{Figures/sim2.pdf}}%
    \put(0.06199042,0.60178547){\color[rgb]{0,0,0}\makebox(0,0)[lt]{\lineheight{1.25}\smash{\begin{tabular}[t]{l}1.0\end{tabular}}}}%
    \put(0.06274707,0.11968129){\color[rgb]{0,0,0}\makebox(0,0)[lt]{\lineheight{1.25}\smash{\begin{tabular}[t]{l}0.0\end{tabular}}}}%
    \put(0.06274609,0.36268584){\color[rgb]{0,0,0}\makebox(0,0)[lt]{\lineheight{1.25}\smash{\begin{tabular}[t]{l}0.5\end{tabular}}}}%
    \put(-0,0.36300842){\color[rgb]{0,0,0}\makebox(0,0)[lt]{\lineheight{1.25}\smash{\begin{tabular}[t]{l}\vect{s}\end{tabular}}}}%
    \put(0.18128194,0.04803463){\color[rgb]{0,0,0}\makebox(0,0)[lt]{\lineheight{1.25}\smash{\begin{tabular}[t]{l}0\end{tabular}}}}%
    \put(0.36758085,0.04825838){\color[rgb]{0,0,0}\makebox(0,0)[lt]{\lineheight{1.25}\smash{\begin{tabular}[t]{l}5\end{tabular}}}}%
    \put(0.54718192,0.04803216){\color[rgb]{0,0,0}\makebox(0,0)[lt]{\lineheight{1.25}\smash{\begin{tabular}[t]{l}10\end{tabular}}}}%
    \put(0.73313904,0.04825838){\color[rgb]{0,0,0}\makebox(0,0)[lt]{\lineheight{1.25}\smash{\begin{tabular}[t]{l}15\end{tabular}}}}%
    \put(0.91924223,0.04803216){\color[rgb]{0,0,0}\makebox(0,0)[lt]{\lineheight{1.25}\smash{\begin{tabular}[t]{l}20\end{tabular}}}}%
    \put(0.55060976,0.0027996){\color[rgb]{0,0,0}\makebox(0,0)[lt]{\lineheight{1.25}\smash{\begin{tabular}[t]{l}\vect{t}\end{tabular}}}}%
  \end{picture}%
\endgroup%

%% file: Figures/v_and_f.pdf_tex
\begingroup%
  \makeatletter%
  \providecommand\color[2][]{%
    \errmessage{(Inkscape) Color is used for the text in Inkscape, but the package 'color.sty' is not loaded}%
    \renewcommand\color[2][]{}%
  }%
  \providecommand\transparent[1]{%
    \errmessage{(Inkscape) Transparency is used (non-zero) for the text in Inkscape, but the package 'transparent.sty' is not loaded}%
    \renewcommand\transparent[1]{}%
  }%
  \providecommand\rotatebox[2]{#2}%
  \newcommand*\fsize{\dimexpr\f@size pt\relax}%
  \newcommand*\lineheight[1]{\fontsize{\fsize}{#1\fsize}\selectfont}%
  \ifx\svgwidth\undefined%
    \setlength{\unitlength}{267.44812252bp}%
    \ifx\svgscale\undefined%
      \relax%
    \else%
      \setlength{\unitlength}{\unitlength * \real{\svgscale}}%
    \fi%
  \else%
    \setlength{\unitlength}{\svgwidth}%
  \fi%
  \global\let\svgwidth\undefined%
  \global\let\svgscale\undefined%
  \makeatother%
  \begin{picture}(1,1.72020402)%
    \lineheight{1}%
    \setlength\tabcolsep{0pt}%
    \put(0,0){\includegraphics[width=\unitlength,page=1]{Figures/v_and_f.pdf}}%
    \put(0.06246676,1.61368372){\color[rgb]{0,0,0}\makebox(0,0)[lt]{\lineheight{1.25}\smash{\begin{tabular}[t]{l}12.0\end{tabular}}}}%
    \put(0.06246676,1.02326068){\color[rgb]{0,0,0}\makebox(0,0)[lt]{\lineheight{1.25}\smash{\begin{tabular}[t]{l}11.0\end{tabular}}}}%
    \put(0.1845184,0.06875035){\color[rgb]{0,0,0}\makebox(0,0)[lt]{\lineheight{1.25}\smash{\begin{tabular}[t]{l}0.0\end{tabular}}}}%
    \put(0.55386545,0.06875051){\color[rgb]{0,0,0}\makebox(0,0)[lt]{\lineheight{1.25}\smash{\begin{tabular}[t]{l}0.5\end{tabular}}}}%
    \put(0.92694082,0.06875051){\color[rgb]{0,0,0}\makebox(0,0)[lt]{\lineheight{1.25}\smash{\begin{tabular}[t]{l}1.0\end{tabular}}}}%
    \put(0.0854066,0.83436738){\color[rgb]{0,0,0}\makebox(0,0)[lt]{\lineheight{1.25}\smash{\begin{tabular}[t]{l}1.0\end{tabular}}}}%
    \put(0.08721374,0.49835357){\color[rgb]{0,0,0}\makebox(0,0)[lt]{\lineheight{1.25}\smash{\begin{tabular}[t]{l}0.5\end{tabular}}}}%
    \put(0.08721374,0.16176095){\color[rgb]{0,0,0}\makebox(0,0)[lt]{\lineheight{1.25}\smash{\begin{tabular}[t]{l}0.0\end{tabular}}}}%
    \put(0.06246676,1.31593052){\color[rgb]{0,0,0}\makebox(0,0)[lt]{\lineheight{1.25}\smash{\begin{tabular}[t]{l}11.5\end{tabular}}}}%
    \put(0.31675222,0.81777302){\color[rgb]{0,0,0}\makebox(0,0)[lt]{\lineheight{1.25}\smash{\begin{tabular}[t]{l}$V_0= 0.10$\\$V_0= 0.13$\\$V_0= 0.15$\end{tabular}}}}%
    \put(0.03124879,0.45698169){\color[rgb]{0,0,0}\rotatebox{90}{\makebox(0,0)[lt]{\lineheight{1.25}\smash{\begin{tabular}[t]{l}$\vect f\left(\vect s,\vect a\right)$\end{tabular}}}}}%
    \put(0.03124879,1.04964429){\color[rgb]{0,0,0}\rotatebox{90}{\makebox(0,0)[lt]{\lineheight{1.25}\smash{\begin{tabular}[t]{l}$E[V^\star(\vect s_+)\,|\, \vect s,\vect a] \text{ and } V^\star(\vect f(\vect s,\vect a)) + V_0$\end{tabular}}}}}%
    \put(0.53656936,0.0067078){\color[rgb]{0,0,0}\makebox(0,0)[lt]{\lineheight{1.25}\smash{\begin{tabular}[t]{l}$\vect{s} +  \vect{a}$\end{tabular}}}}%
  \end{picture}%
\endgroup%

%% file: Draft.bbl
\begin{thebibliography}{10}
\providecommand{\url}[1]{#1}
\csname url@samestyle\endcsname
\providecommand{\newblock}{\relax}
\providecommand{\bibinfo}[2]{#2}
\providecommand{\BIBentrySTDinterwordspacing}{\spaceskip=0pt\relax}
\providecommand{\BIBentryALTinterwordstretchfactor}{4}
\providecommand{\BIBentryALTinterwordspacing}{\spaceskip=\fontdimen2\font plus
\BIBentryALTinterwordstretchfactor\fontdimen3\font minus \fontdimen4\font\relax}
\providecommand{\BIBforeignlanguage}[2]{{%
\expandafter\ifx\csname l@#1\endcsname\relax
\typeout{** WARNING: IEEEtran.bst: No hyphenation pattern has been}%
\typeout{** loaded for the language `#1'. Using the pattern for}%
\typeout{** the default language instead.}%
\else
\language=\csname l@#1\endcsname
\fi
#2}}
\providecommand{\BIBdecl}{\relax}
\BIBdecl

\bibitem{puterman2014markov}
M.~L. Puterman, \emph{Markov decision processes: discrete stochastic dynamic programming}.\hskip 1em plus 0.5em minus 0.4em\relax John Wiley \& Sons, 2014.

\bibitem{van2012reinforcement}
M.~Van~Otterlo and M.~Wiering, ``Reinforcement learning and markov decision processes,'' in \emph{Reinforcement learning: State-of-the-art}.\hskip 1em plus 0.5em minus 0.4em\relax Springer, 2012, pp. 3--42.

\bibitem{feinberg2012handbook}
E.~A. Feinberg and A.~Shwartz, \emph{Handbook of Markov decision processes: methods and applications}.\hskip 1em plus 0.5em minus 0.4em\relax Springer Science \& Business Media, 2012, vol.~40.

\bibitem{MPCbook}
J.~B. Rawlings, D.~Q. Mayne, and M.~Diehl, \emph{Model predictive control: theory, computation, and design}.\hskip 1em plus 0.5em minus 0.4em\relax Nob Hill Publishing Madison, WI, 2017, vol.~2.

\bibitem{bertsekas2019reinforcement}
D.~Bertsekas, \emph{Reinforcement learning and optimal control}.\hskip 1em plus 0.5em minus 0.4em\relax Athena Scientific, 2019.

\bibitem{rawlings2017model}
J.~B. Rawlings, D.~Q. Mayne, and M.~Diehl, \emph{Model predictive control: theory, computation, and design}.\hskip 1em plus 0.5em minus 0.4em\relax Nob Hill Publishing Madison, WI, 2017, vol.~2.

\bibitem{bemporad2007robust}
A.~Bemporad and M.~Morari, ``Robust model predictive control: A survey,'' in \emph{Robustness in identification and control}.\hskip 1em plus 0.5em minus 0.4em\relax Springer, 2007, pp. 207--226.

\bibitem{mesbah2016stochastic}
A.~Mesbah, ``Stochastic model predictive control: An overview and perspectives for future research,'' \emph{IEEE Control Systems Magazine}, vol.~36, no.~6, pp. 30--44, 2016.

\bibitem{langson2004robust}
W.~Langson, I.~Chryssochoos, S.~V. Rakovi{\'c}, and D.~Q. Mayne, ``Robust model predictive control using tubes,'' \emph{Automatica}, vol.~40, no.~1, pp. 125--133, Jan. 2004.

\bibitem{kordabad2023equivalence}
A.~B. Kordabad, M.~Zanon, and S.~Gros, ``Equivalence of optimality criteria for markov decision process and model predictive control,'' \emph{IEEE Transactions on Automatic Control}, 2023.

\bibitem{gros2022economic}
S.~Gros and M.~Zanon, ``Economic mpc of markov decision processes: Dissipativity in undiscounted infinite-horizon optimal control,'' \emph{Automatica}, vol. 146, p. 110602, 2022.

\bibitem{gros2020linear}
S.~Gros, M.~Zanon, R.~Quirynen, A.~Bemporad, and M.~Diehl, ``From linear to nonlinear mpc: bridging the gap via the real-time iteration,'' \emph{International Journal of Control}, vol.~93, no.~1, pp. 62--80, 2020.

\bibitem{dockner1991optimality}
E.~J. Dockner and G.~Feichtinger, ``On the optimality of limit cycles in dynamic economic systems,'' \emph{Journal of Economics}, vol.~53, no.~1, pp. 31--50, 1991.

\bibitem{amrit2011economic}
R.~Amrit, J.~B. Rawlings, and D.~Angeli, ``Economic optimization using model predictive control with a terminal cost,'' \emph{Annual Reviews in Control}, vol.~35, no.~2, pp. 178--186, Dec. 2011.

\bibitem{gros2019data}
S.~Gros and M.~Zanon, ``Data-driven economic nmpc using reinforcement learning,'' \emph{IEEE Transactions on Automatic Control}, vol.~65, no.~2, pp. 636--648, 2019.

\bibitem{diehl2011lyapunov}
M.~Diehl, R.~Amrit, and J.~B. Rawlings, ``A lyapunov function for economic optimizing model predictive control,'' \emph{IEEE Transactions on Automatic Control}, vol.~56, no.~3, pp. 703--707, 2011.

\bibitem{zanon2022new}
M.~Zanon and S.~Gros, ``A new dissipativity condition for asymptotic stability of discounted economic mpc,'' \emph{Automatica}, vol. 141, p. 110287, 2022.

\bibitem{zanon2016tracking}
M.~Zanon, S.~Gros, and M.~Diehl, ``A tracking {{MPC}} formulation that is locally equivalent to economic {{MPC}},'' \emph{Journal of Process Control}, vol.~45, pp. 30--42, Sep. 2016.

\bibitem{muller2015necessity}
M.~A. M{\"u}ller, D.~Angeli, and F.~Allg{\"o}wer, ``On {{Necessity}} and {{Robustness}} of {{Dissipativity}} in {{Economic Model Predictive Control}},'' \emph{IEEE Transactions on Automatic Control}, vol.~60, no.~6, pp. 1671--1676, Jun. 2015.

\bibitem{ljung1999system}
\BIBentryALTinterwordspacing
L.~Ljung, \emph{System Identification: Theory for the User}, ser. Prentice Hall information and system sciences series.\hskip 1em plus 0.5em minus 0.4em\relax Prentice Hall PTR, 1999. [Online]. Available: \url{https://books.google.no/books?id=nHFoQgAACAAJ}
\BIBentrySTDinterwordspacing

\end{thebibliography}
